\documentclass[sigconf,nonacm]{acmart}

\usepackage{listings}
\usepackage{graphicx} 
\usepackage{amsmath}
\usepackage{amsfonts}
\usepackage{algorithm}
\usepackage{algorithmic}
\usepackage{etoolbox} 
\usepackage{multirow}
\usepackage{tablefootnote}
\usepackage{makecell}
\usepackage{enumitem}
\usepackage{subcaption}
\usepackage{setspace}
\usepackage{wrapfig}
\usepackage[most]{tcolorbox}
\usepackage[framemethod=tikz]{mdframed}
\usepackage{hyperref}

\AtBeginDocument{%
  }

\setcopyright{acmlicensed}
\copyrightyear{2018}
\acmYear{2018}
\acmDOI{XXXXXXX.XXXXXXX}

\definecolor{gray}{HTML}{808080}
\definecolor{teal}{HTML}{21908C}
\definecolor{yellow}{HTML}{FDE725}
\definecolor{blue}{HTML}{1E2BC5}
\definecolor{purple}{HTML}{440154}
\definecolor{green}{HTML}{238E12}

\newtoggle{comments}
\toggletrue{comments}

\iftoggle{comments}{
    \newcommand{\todo}[1] {{\color{red}[{\bf TODO:} #1]}}
    \newcommand{\cut}[1] {{\color{red}\sout{#1}}}
    \newcommand{\notezg}[1] {{\color{teal}[{\bf ZG:} #1]}}
    \newcommand{\notejz}[1] {{\color{green}[{\bf JZ:} #1]}}
    \newcommand{\notetj}[1] {{\color{blue}[{\bf Thorsten:} #1]}}

}{
    \newcommand{\todo}[1] {}
    \newcommand{\cut}[1] {}
    \newcommand{\notezg}[1] {}
    \newcommand{\notejz}[1] {}
    \newcommand{\notetj}[1] {}

}

\newcommand{\custompipelinename} {\textsc{LangPTune}}
\newcommand{\encoder}{{\pi}}     
\newcommand{\decoder}{{\sigma}}  
\newcommand{\embedding}{\Phi}    
\newcommand{\similarity}{{sim}}  
\newcommand{\ranking}{r}
\newcommand{\util}{U}            
\newcommand{\profile}{p}
\newcommand{\history}{h}
\newcommand{\future}{f}
\newcommand{\ritem}{i}

\newcommand{\descript}{d}        
\newcommand{\metadata}{D}

\DeclareMathOperator*{\argmax}{arg\,max}
\DeclareMathOperator*{\argmin}{arg\,min}
\DeclareMathOperator*{\argsort}{arg\,sort}

\gdef\Sepline{%
  \par\noindent\makebox[\linewidth][l]{%
  \hspace*{-\mdflength{innerleftmargin}}%
   \tikz\draw[thick,dashed,gray!60] (0,0) --%
        (\textwidth+\the\mdflength{innerleftmargin}+\the\mdflength{innerrightmargin},0);
  }\par\nobreak}

\acmConference[Conference acronym 'XX]{Make sure to enter the correct
  conference title from your rights confirmation emai}{June 03--05,
  2018}{Woodstock, NY}
\acmISBN{978-1-4503-XXXX-X/18/06}

\begin{document}

\definecolor{perfblue}{RGB}{64, 114, 175}

\title{End-to-end Training for Recommendation with \\ Language-based User Profiles}


\author{Zhaolin Gao}
\affiliation{%
  \institution{Cornell University}
  \city{Ithaca}
  \country{USA}}
\email{zg292@cornell.edu}

\author{Joyce Zhou}
\affiliation{%
  \institution{Cornell University}
  \city{Ithaca}
  \country{USA}}
\email{jz549@cornell.edu}

\author{Yijia Dai}
\affiliation{%
  \institution{Cornell University}
  \city{Ithaca}
  \country{USA}}
\email{yd73@cornell.edu}

\author{Thorsten Joachims}
\affiliation{%
  \institution{Cornell University}
  \city{Ithaca}
  \country{USA}}
\email{tj36@cornell.edu}









\begin{abstract}
There is a growing interest in natural language-based user profiles for recommender systems, which aims to enhance transparency and scrutability compared with embedding-based methods. Existing studies primarily generate these profiles using zero-shot inference from large language models (LLMs), but their quality remains insufficient, leading to suboptimal recommendation performance. In this paper, we introduce \custompipelinename, the first end-to-end training framework to optimize LLM-generated user profiles. Our method significantly outperforms zero-shot approaches by explicitly training the LLM for the recommendation objective. Through extensive evaluations across diverse training configurations and benchmarks, we demonstrate that \custompipelinename\ not only surpasses zero-shot baselines but can also matches the performance of state-of-the-art embedding-based methods. Finally, we investigate whether the training procedure preserves the interpretability of these profiles compared to zero-shot inference through both GPT-4 simulations and crowdworker user studies. Implementation of \custompipelinename\ can be found at \href{https://github.com/ZhaolinGao/LangPTune}{\color{perfblue}https://github.com/ZhaolinGao/LangPTune}.
\end{abstract}



\begin{teaserfigure}
  \centering
  \includegraphics[scale=0.55,trim={20 190 20 180},,clip]{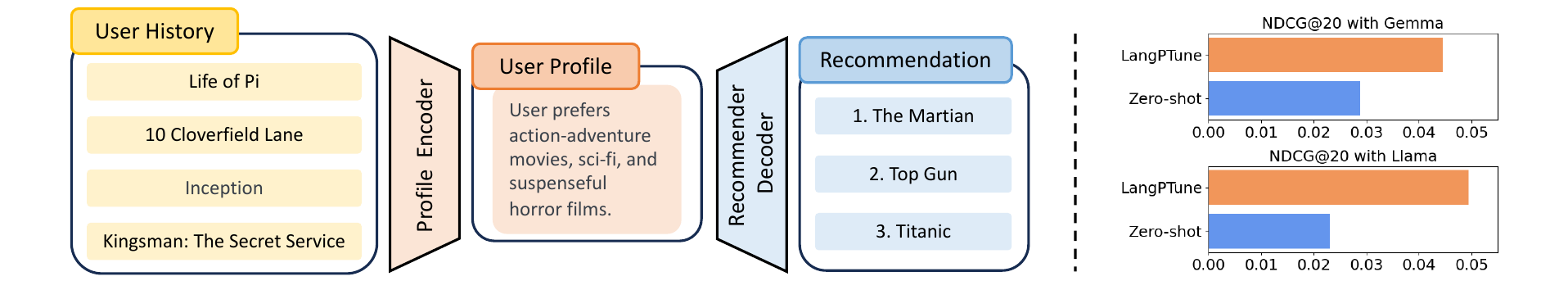}
  \vskip -0.4cm
  \caption{(Left) An illustration of \custompipelinename\ inference pipeline. The profile encoder first generates a language-based user profile, and the recommender decoder recommends based on the generated profile. (Right) Evaluation results on the Amazon-Movie-TV dataset show that \custompipelinename\ significantly outperforms zero-shot method on both Llama and Gemma models.
  \Description[]{}
  \label{fig:pipeline}}
\end{teaserfigure}



\maketitle

\section{Introduction}

Accurate recommendations have become an essential part of web platforms such as multimedia streaming services, e-commerce, and social media. To power these systems, recommender system research has focused on learning embeddings of users and items to represent user preferences~\citep{hu2008colla,5197422,NGCF,SRNS, he2020lightgcnsimplifyingpoweringgraph, gao2022mcl, yang2024notall}. While these embeddings are computationally efficient, their inherently high dimensionality makes them difficult to interpret. This limitation poses a transparency barrier, making it difficult for the user to understand, modify, and steer the recommendations.

To overcome the fundamental transparency limitations of conventional recommender systems, previous studies have proposed and explored the use of language-based user profiles, where the user's preferences are represented as human-readable free-form text rather than dense vectors~\cite{Radlinski_2022, yang2023palrpersonalizationawarellms, ramos2024transparentscrutablerecommendationsusing, zhou2024languagebaseduserprofilesrecommendation, penalozaTEARSTextualRepresentations2024}. Such language-based user profiles are inherently more interpretable than vector-based profiles~\citep{gengRecommendationLanguageProcessing2023, lazarMoralCaseUsing2024}, and they promise improved transparency, scrutability, and ultimately steerability of recommendation platforms~\citep{liu2023chatgptgoodrecommenderpreliminary, sanner_coldstart}. Furthermore, language-based profiles can provide additional features to traditional recommender systems, boosting their performance and enabling more creative personalization affordances~\citep{Ren_2024, lyu2024llmrecpersonalizedrecommendationprompting, yang2024sequentialrecommendationlatentrelations, penalozaTEARSTextualRepresentations2024}. 

Despite these advantages, prior work has not addressed how to train large language models (LLMs) to generate effective user profiles optimized for downstream recommendation. Existing approaches primarily rely on zero-shot or few-shot inference~\cite{Radlinski_2022, yang2023palrpersonalizationawarellms, zhou2024languagebaseduserprofilesrecommendation} or prompt tuning~\cite{ramos2024transparentscrutablerecommendationsusing}. In this paper, we introduce \textbf{the first end-to-end training pipeline that optimizes LLMs to generate user profiles for the recommendation objective}. The inference process of our pipeline is illustrated in Fig.~\ref{fig:pipeline}. Given a list of items that the user has interacted with, our goal is to produce a high-quality user profile via a profile encoder (i.e. LLM) that maximizes the downstream performance of the recommender decoder. Drawing on insights from Reinforcement Learning from Human Feedback (RLHF)~\cite{Christiano2017rlhf,ziegler2020finetuninglanguagemodelshuman}, we propose a novel approach called \textit{Reinforcement Learning for System Optimization} (\textbf{RLSO}) that optimizes the LLM based on the feedback from the recommendation system. For each user, we sample multiple profiles from the LLM and use the performance of the downstream recommender as reward signals to optimize the LLM, as illustrated in Fig.~\ref{fig:llm_optimization}.
Furthermore, we jointly optimize the downstream ranking model of the recommender alongside the LLM through Contrastive Learning (CL)~\citep{chen2020simpleframeworkcontrastivelearning}, which completes our end-to-end training framework that we call \textbf{\custompipelinename} (\textit{Lang}uage-based \textit{P}rofile \textit{Tune}r). 

We implement our framework using Mxbai~\cite{emb2024mxbai} as the ranking model in the recommender decoder and evaluate the effectiveness of our algorithm on two different LLMs as profile encoder, Gemma-2B-it~\citep{gemmateam2024gemma2improvingopen} and Llama-3-8B-it~\citep{dubey2024llama3herdmodels}. We compare \custompipelinename\ against a variety of baseline methods spanning different recommendation approaches on multiple public datasets. Our end-to-end training significantly outperforms other language-based profile methods and often exhibits performance comparable to the conventional embedding-based recommendation models. The major contributions of this paper are summarized below:
\begin{itemize}[leftmargin=*,noitemsep,topsep=0pt]
  \item We introduce \custompipelinename, the first end-to-end training pipeline for effective generation of user profiles, with detailed theoretical derivations and intuitive explanations of our algorithm.
  \item We implement our framework using state-of-the-art LLMs and ranking models, demonstrating that it significantly outperforms other interpretable profile methods and exhibits performance comparable to conventional recommendation models.
  \item We study the effects of profile length and training dataset size, and provide a qualitative analysis of the generated user profiles.
  \item We conduct studies with both human participants and GPT-4 to compare the interpretability of the trained LLM against a zero-shot LLM, ensuring that interpretability is not compromised.
\end{itemize}

\section{Related Work}

\textbf{LLM for Recommender Systems.} Our research falls within the broad spectrum of using LLMs for recommendation, which can largely be categorized into two main approaches. The first approach involves directly training LLMs for item or rating prediction~\citep{zhang2023collmintegratingcollaborativeembeddings, ji2023genreclargelanguagemodel, Bao_2023, kang2023llmsunderstanduserpreferences, liao2024llaralargelanguagerecommendationassistant, gengRecommendationLanguageProcessing2023, xuPromptingLargeLanguage2024a}, or otherwise using them as system augmentation~\citep{liCTRLConnectCollaborative2023, xiOpenWorldRecommendationKnowledge2023a}. With this approach, user interaction histories and item metadata are provided as inputs to the LLM, which is then trained to predict the items most relevant to the user. These works take advantage of LLM's ability to learn complex patterns for item recommendation, and generally do not prioritize transparency or explainability. The second approach leverages LLMs to generate additional user or item profiles/descriptions based on existing features in a zero-shot manner~\cite{Harte_2023, Ren_2024, lyu2024llmrecpersonalizedrecommendationprompting, yang2024sequentialrecommendationlatentrelations}. These works demonstrate that incorporating additional features generated by LLMs could enhance the performance of conventional recommendation systems. Our work aligns with this second approach, but rather than relying solely on zero-shot inference, we aim to directly optimize the LLM to generate more effective profiles for recommendation.

\textbf{Natural language-based profile methods.}
We focus on centering recommender systems around natural language-based user profiles, which has garnered increasing attention for enabling increased transparency, explainability, and scrutability of recommendation systems~\citep{Radlinski_2022}.
In comparison with previous work based on user interest concept sets~\citep{mysoreEditableUserProfiles2023, balogTransparentScrutableExplainable2019}, natural language profiles offer increased nuance, flexibility, and affordances for user control~\citep{gengRecommendationLanguageProcessing2023, Radlinski_2022, lazarMoralCaseUsing2024}.
There have been a number of works on generating initial profiles based on user interaction history, as well as using profiles to make recommendations. 
For example, \citet{yang2023palrpersonalizationawarellms} constructs user profiles based on observed user behaviors and uses them to prompt an LLM to perform personalized recommendations. \citet{ramos2024transparentscrutablerecommendationsusing} explores various prompting strategies aimed at improving the quality of LLM-generated user profiles. Their work also examines the system's scrutability by observing how modifications to the user profile affect the resulting recommendations. \citet{zhou2024languagebaseduserprofilesrecommendation} demonstrates that generating a compact and human-readable summary often performs comparably with or better than direct LLM prediction.
However, none of these approaches allow for end-to-end training to optimize profile generation.

\textbf{Reinforcement Learning from Human Feedback (RLHF).} Our method of training LLMs is closely related to the families of techniques for preference fine-tuning. One approach involves first training a reward model on a dataset of human preferences, which is then used to provide rewards to a downstream reinforcement learning algorithm (e.g., PPO~\citep{schulman2017proximalpolicyoptimizationalgorithms, ziegler2020finetuninglanguagemodelshuman}). LLMs fine-tuned using this approach include GPT-N \citep{gpt4}, Claude-N \citep{claude3}, and Llama-N \citep{llama3}. Another approach directly optimizes the LLM using offline human preference data without fitting a reward model. This family of methods includes DPO \citep{rafailov2023direct}, IPO \citep{azar2023generaltheoreticalparadigmunderstand}, and KTO \citep{ethayarajh2024ktomodelalignmentprospect}, which are simpler to implement due to the removal of the reward model. Our approach is an iterative variant of the latter approach, where we employ an offline algorithm on iteratively collected profiles, with rewards generated by the recommender decoder.

\section{\custompipelinename: \textit{Lang}uage-based \textit{P}rofile \textit{Tune}r}

In this section, we formalize our encoder-decoder recommendation model and derive our end-to-end training framework.

\subsection{\custompipelinename\ Model Formulation}

The inference process of our pipeline is illustrated in Fig.~\ref{fig:pipeline}. Given a list of items that the user has interacted with, our goal is to generate a high-quality user profile using a profile encoder that maximizes the downstream performance of the recommender decoder.

\subsubsection{Recommender Decoder $\decoder$.} The goal for the recommender decoder is to provide accurate recommendation based on the input profile $\profile$. Specifically, given a set of $I$ items, the recommender decoder $\decoder$ generates a ranked list of $I$ items based on the input profile $\profile$ to maximize the ranking quality $\util$. Let $\ritem \in \{\ritem_1, \ldots, \ritem_{I}\}$ denote an item, and each item $\ritem$ has associated metadata $\descript_\ritem \in \mathcal{D}$ such as title, description, and category.
Given a user profile $\profile$ and a complete set of item metadata $\mathcal{D}$, the decoder $\decoder$ computes a similarity score $\similarity(\profile,\descript_\ritem)$ for each item $\ritem$ and generates a ranked list $\ranking$ via sorting.
\begin{equation}
    \ranking = \decoder(\profile,\mathcal{D}) = \argsort_{\descript_i \in \mathcal{D}} \left[ \similarity(\profile,\descript_\ritem) \right]
\end{equation}
The similarity model $\similarity(\profile,\descript_\ritem)$ can take a wide range of forms, but for simplicity we focus on a model that learns a semantic embedding $\embedding$ of both profiles and items, and then computes similarity via the inner product 
\begin{equation}
\similarity(\profile,\descript_\ritem)=\embedding(\profile) \cdot \embedding(\descript_\ritem).
\end{equation}
However, there is a wide range of alternative similarity models (e.g. two-tower models \citep{multiviewali2015,covinton2016youtube}) that can equally be trained with the method we will describe in Section~\ref{sec:traindecoder}. Let $\future = [\ritem_{1}, \ldots, \ritem_{|\future|}]$ denote the list of items the user will interact with in the future. The ranking quality is then evaluated based on the ranking $\ranking$ for test items in $\future$, denoted as $\util(\ranking, \future)$. This evaluation is typically performed using ranking metrics such as Mean Reciprocal Rank (MRR) or Normalized Discounted Cumulative Gain (NDCG).

A key advantage of our similarity-based decoder is its ability to scale to large collections of items, since the $\argsort$ can be efficiently implemented with conventional ranking techniques. This overcomes the bottleneck of previous works~\citep{yang2023palrpersonalizationawarellms, zhou2024languagebaseduserprofilesrecommendation} that typically use an LLM as the recommender. While an LLM can directly operate on the textual profiles without further training, using zero-shot LLM-based recommenders typically implies long latency and it requires limiting to small subsets of all items due to the context length restrictions of LLMs. In contrast, our system is designed to scale across large item spaces while incorporating textual input and leveraging natural language user profiles.

\subsubsection{Profile Encoder $\encoder$.} The profile encoder $\pi$ computes a mapping from the user's interaction history to a natural-language profile. Let $\history = [\ritem_1, \ritem_2, \ldots, \ritem_{|\history|}]$ represent a particular user's interaction history, consisting of a sequence of items, and denote the interaction history annotated with the metadata of the items as $\metadata(\history) = [\descript_{\ritem_1}, \ldots, \descript_{\ritem_{|\history|}}]$.
We model the profile encoder as a distribution over the profile space $\Delta \mathcal{P}$ from which we can efficiently generate a profile $\profile$ by sampling.
\begin{equation}
    \profile \sim \encoder(\cdot|\metadata(\history))
\end{equation}

A natural question may rise regarding the necessity of utilizing a profile encoder rather than directly feeding the the metadata of items in the user interaction history, $\metadata(\history)$, to the decoder. Since the generated profile is a higher abstraction than a raw list of items, it can encapsulate more generalizable preferences and interests instead of merely reflecting specific past items. In addition, the profiles are more compact than the metadata, which enhances the interpretability of the profile and the steerability for users to modify or expand their profiles.

\subsubsection{\custompipelinename.} Combining the profile encoder $\encoder$ and the recommender decoder $\decoder$, we arrive at our proposed framework \custompipelinename. Given a user with history $\history$, future interactions $\future$, and item metadata $\mathcal{D}$, the encoder first generates a user profile $\profile$ based on the metadata of items that are in the user interaction history, $\profile \sim \encoder(\cdot | \metadata(\history))$. Then, the decoder generates a ranked list given the profile and item metadata, $\ranking = \decoder(\profile, \mathcal{D})$. The framework's objective is to maximize recommendation performance $\util(\ranking, \future)$ while ensuring that the user profile $\profile$ is expressed in a well-formed natural language format.







\subsection{\custompipelinename\ Training Algorithm} 

\begin{figure}[t]
    \centering
    \includegraphics[scale=0.6,trim={270 140 270 170},clip]{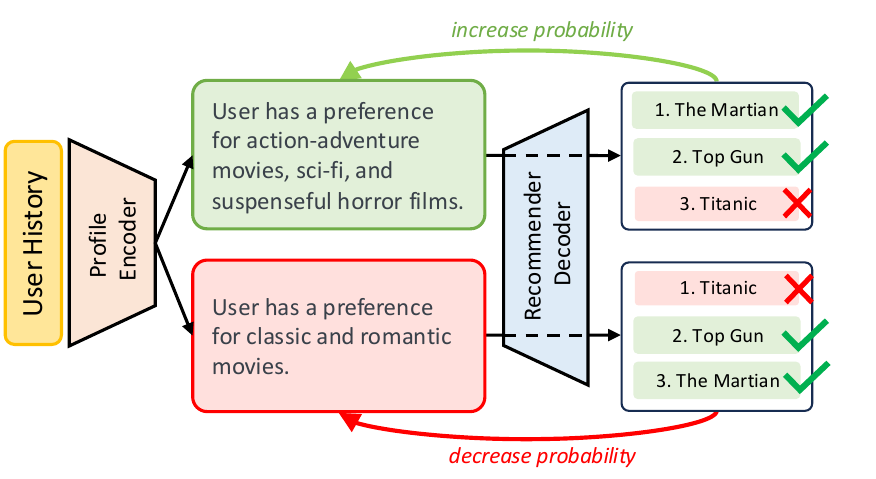}
    \vskip -0.4cm
    \caption{\textbf{Reinforcement Learning for System Optimization Pipeline} \label{fig:llm_optimization}}
    \Description[]{}
    \vskip -0.3cm
\end{figure}

In the section, we first introduce our algorithms for training the profile encoder $\encoder$ and the recommender decoder $\decoder$ separately. We then integrate both methods into a unified end-to-end training framework.

\subsubsection{RLSO for Profile Encoder Optimization} \label{sec:rlso}

In this section, we derive our training algorithm for profile encoder optimization which we refer to as \textit{Reinforcement Learning for System Optimization} (\textbf{RLSO}) as illustrated in Fig.~\ref{fig:llm_optimization}. To optimize the encoder $\pi$ such that it maximizes the performance of the recommender decoder measured by $\util$, we aim to solve the following Reinforcement Learning (RL) problem constrained by Kullback–Leibler (KL) divergence at each iteration of our algorithm:
\begin{align}
    \encoder_{t+1} = \argmax_\encoder \ 
 &\mathbb{E}_{\history, \future, \profile \sim \encoder(\cdot | \metadata(\history))} \biggl[ \util(\decoder(\profile, \mathcal{D}), \future) - \nonumber \\
    &\frac{1}{\eta}\text{KL}\left( \encoder(\cdot | \metadata(\history)) || \encoder_t(\cdot | \metadata(\history)) \right) \biggr]
    \label{eq:linear_KL_md}
\end{align}
where $\eta$ is a parameter that controls the degree of KL regularization, and $\pi_t$ is the encoder from previous iteration. Intuitively, the objective is to maximize the metric of the recommender system, $\util(\decoder(\profile, \mathcal{D}), \future)$, while staying close to $\encoder_t$ by minimizing the KL divergence between $\encoder$ and $\encoder_t$. The KL divergence is typically used in RL to prevent reward hacking. In this context, it serves to preserve the readability of the user profiles by staying close to the encoder $\pi_t$ of the previous iteration.

Now we start to derive the loss function for the above objective. From \citet{ziebart2008maximum}, there exists a closed-form solution to the above minimum relative entropy problem (Eq.~\ref{eq:linear_KL_md})~\citep{Gr_nwald_2004}:
\begin{align}
\label{eq:md}
\forall \history, \future, \profile:\ &\encoder_{t+1}(\profile|\metadata(\history)) \nonumber 
=\frac{ \encoder_t(\profile|\metadata(\history)) \exp(\eta \util(\decoder(\profile, \mathcal{D}), \future))}{ Z(\history, \future)} \\
&Z(\history, \future) = \sum_{\profile} \encoder_t(\profile|\metadata(\history)) \exp(\eta \util(\decoder(\profile, \mathcal{D}), \future))
\end{align}
where $Z(\history, \future)$ is a partition function to ensure that $\encoder_{t+1}(\cdot|\metadata(\history))$ remains as a valid probability distribution that sums to 1.
Based on \citet{rafailov2023direct}, we can invert Eq. \ref{eq:md} and write the ranking quality as a function of $\pi$:
\begin{equation}
\label{eq:dpo}
\resizebox{1\hsize}{!}{$\forall \history, \future, \profile:   \util(\decoder(\profile, \mathcal{D}), \future) = \frac{1}{\eta} \left( \ln(Z(\history, \future)) + \ln \left (\frac{\encoder_{t+1}(\profile|\metadata(\history))}{\encoder_t(\profile|\metadata(\history))} \right) \right)$}
\end{equation}
Note that the partition function $Z(\history, \future)$ does not depend on profile $\profile$. Therefore, we can sample another profile $\profile' \sim \encoder(\cdot | \metadata(\history))$ and take the difference of the above expression across two profiles to eliminate the partition function:
\begin{align}
\label{eq:dpo_no_partition}
\forall \history, \future, \profile, \profile': \  &\util(\decoder(\profile, \mathcal{D}), \future) - \util(\decoder(\profile', \mathcal{D}), \future) = \nonumber \\
&\frac{1}{\eta} \left( \ln \left (\frac{\encoder_{t+1}(\profile|\metadata(\history))}{\encoder_t(\profile|\metadata(\history))} \right)  - \ln \left (\frac{\encoder_{t+1}(\profile'|\metadata(\history))}{\encoder_t(\profile'|\metadata(\history))} \right)\right)
\end{align}
Following \citet{gao2024rebel}, Eq.~\ref{eq:dpo_no_partition} can be directly optimized with a squared loss objective:
\begin{align}
\Biggl(&\util(\decoder(\profile, \mathcal{D}), \future) - \util(\decoder(\profile', \mathcal{D}), \future) - \nonumber \\
&\frac{1}{\eta} \left( \ln \left (\frac{\encoder_{t+1}(\profile|\metadata(\history))}{\encoder_t(\profile|\metadata(\history))} \right)  - \ln \left (\frac{\encoder_{t+1}(\profile'|\metadata(\history))}{\encoder_t(\profile'|\metadata(\history))} \right)\right) \Biggr)^2
\label{eq:rlso}
\end{align}

\begin{algorithm}[t]
    \caption{\textbf{\custompipelinename} Optimization}
    \begin{algorithmic}[1]
        \STATE \textbf{Required}: ranking quality function $\util$, 
        parameter $\eta$, 
        similarity function $sim(\cdot, \cdot)$
        \STATE Initialize encoder $\encoder$ and decoder $\decoder$ which uses $\embedding$ as its embedding model.
        \FOR{$k = 0$ to $K-1$}
            \STATE $\embedding_0 \leftarrow \embedding$
            \FOR{$j = 0$ to $J-1$}
                \STATE Collect data $\epsilon_j = \{\history, \profile, B\}$ where $\profile \sim \encoder(\cdot | \metadata(\history))$, $B$ is a list of items with one positive item $\ritem_+$.
                \STATE Optimize for Eq.~\ref{eq:infonce} to obtain $\embedding_{j+1}$.
            \ENDFOR
            \STATE $\embedding \leftarrow \embedding_J$
            \STATE $\encoder_0 \leftarrow \encoder$
            \FOR{$t = 0$ to $T-1$}
                \STATE Collect data $\xi_t = \{\history, \future, \profile, \profile' \}$ where $\profile, \profile' \sim \encoder_t(\cdot | \metadata(\history))$.
                \STATE Solve the squared regression in Eq.~\ref{eq:rlso} to obtain $\encoder_{t+1}$.
            \ENDFOR
            \STATE $\encoder \leftarrow \encoder_{T}$
        \ENDFOR
    \end{algorithmic}
\label{alg:joint}
\end{algorithm}

To optimize the profile encoder with the above squared loss objective given an interaction history $\history$, we begin by sampling two profiles, $\profile \sim \encoder_t(\cdot | \metadata(\history))$ and $ p' \sim \encoder_t(\cdot | \metadata(\history))$, from $\encoder_t$.
Next, we obtain two ranked lists based on the sampled profiles, $\ranking = \decoder(\profile,\mathcal{D})$ and $\ranking' = \decoder(\profile',\mathcal{D})$, and calculate the corresponding ranking qualities, $\util(\ranking, \future)$ and $\util(\ranking', \future)$. The optimization of $\encoder_t$ is illustrated in Fig.~\ref{fig:llm_optimization} which is guided by the relative difference between these ranking qualities. Intuitively, if profile $\profile$ is better than $\profile'$, i.e. $\util(\decoder(\profile, \mathcal{D}), \future) > \util(\decoder(\profile', \mathcal{D}), \future)$, we aim to optimize the encoder to increase the likelihood of generating $\profile$ over $\profile'$, i.e. increasing $\ln \left(\frac{\encoder_{t+1}(\profile|\metadata(\history))}{\encoder_t(\profile|\metadata(\history))} \right) - \ln \left(\frac{\encoder_{t+1}(\profile'|\metadata(\history)}{\encoder_t(\profile'|\metadata(\history))} \right)$. We refer to this training algorithm as \textit{Reinforcement Learning for System Optimization} (\textbf{RLSO}), since it trains the encoder to produce profiles that maximize the performance of the recommendation \textit{system} (i.e. the decoder $\decoder$).

\begin{figure*}[!ht]
\begin{subfigure}{\linewidth}
\begin{tcolorbox}[colback=blue!5!white,colframe=black]
\begin{small}
\textbf{User Interation History:}
Life of Pi, 10 Cloverfield Lane, Inception, Kingsman: The Secret Service
\Sepline
\textbf{Zero-shot Generation:} \newline
\textbf{story is important}: the user appreciates a good story, as evident from their reviews of "" life of pi "" and "" 10 cloverfield lane "". they mention the importance of a great story, message, and script.\newline
\textbf{action and suspense}: the user enjoys action - packed and suspenseful movies, as seen in their reviews of "" 10 cloverfield lane "", "" inception "", and "" kingsman : the secret service "". they appreciate films that keep them engaged and guessing until the end. \newline
\textbf{visually impressive}: the user is impressed by visually stunning movies, as evident from their review of "" life of pi "", which they praised for its great visuals.
\Sepline
\textbf{\custompipelinename\ Generation:} \newline
\textbf{Storytelling}: The user enjoys complex storylines with \textcolor{orange}{twists and surprises}. They appreciate movies that explore themes of \textcolor{orange}{adventure, discovery, and suspense}. The user particularly liked the unexpected connections and surprises in ``Life of Pi'' and the twists in ``10 Cloverfield Lane''.\newline
\textbf{Suspense and Thrills}: The user enjoys suspenseful movies that keep them guessing until the end. They highly rated ``10 Cloverfield Lane'' for its ability to maintain tension and uncertainty throughout the film.\newline
\textbf{Visuals}: The user appreciates visually stunning movies with groundbreaking effects. They praised the visuals in ``Life of Pi'' and \textcolor{orange}{might have appreciated the action sequences and spy-thriller elements in ``Kingsman: The Secret Service''}.
\end{small}
\end{tcolorbox}
\end{subfigure}\hfill
\vspace{-1em}
\caption{Example of the first three points from the profiles generated by Llama-3-8B-it before and after trained by \textbf{\custompipelinename}. The highlighted text represents the additional details generated after training.}
\vspace{-0em}
\label{fig:example_data_and_profile}
\end{figure*}

\textbf{Intuitive Interpretation of RLSO.} From the above derivations, solving Eq.~\ref{eq:rlso} optimally would recover the optimal solution in Eq.~\ref{eq:dpo_no_partition}.
Aggregate over $\profile'$ implies that there must exists a $\profile$-independent function $c(\history, \future)$ such that:
\begin{align}
\forall \history, \future, \profile: \ \frac{1}{\eta} \ln \frac{\encoder_{t+1}(\profile|\metadata(\history))}{\encoder_t(\profile|\metadata(\history))} = \util(\decoder(\profile, \mathcal{D}), \future) + c(\history, \future)
\end{align}
Rearranging the terms, we have:
\begin{align}
\forall \history, \future, \profile: \ &\encoder_{t+1}(\profile|\metadata(\history)) \nonumber\\
&= \encoder_t(\profile|\metadata(\history)) \exp (\eta \util(\decoder(\profile, \mathcal{D}), \future) + \eta c(\history, \future)) \nonumber\\
&\propto \encoder_t(\profile|\metadata(\history)) \exp (\eta \util(\decoder(\profile, \mathcal{D}), \future)) \nonumber
\end{align}
Intuitively, the probability of generating a profile at the next iteration, $\encoder_{t+1}(\profile|\metadata(\history))$, is proportional to the performance of the profile on the recommender decoder, $\exp (\eta \util(\decoder(\profile, \mathcal{D}), \future))$. Better recommendation performance means the encoder is more likely to generate that profile at the next iteration. Note that the above update procedure also recovers the Natural Policy Gradient update with the softmax parametrization~\citep{agarwal2020theorypolicygradientmethods}, which has a fast convergence rate of $O(1/T)$.

\subsubsection{CL for Recommender Decoder Optimization}
\label{sec:traindecoder}

In our implementation, the text-based recommender decoder $\decoder$ contains a text embedding model denoted by $\embedding$. This embedding model is optimized using the standard Contrastive Learning (\textbf{CL}) framework \cite{chen2020simpleframeworkcontrastivelearning}, leveraging the InfoNCE loss \citep{oord2019representationlearningcontrastivepredictive} to enhance its performance. 
Given a user profile $\profile$, a batch of items $B$ with one positive item $\ritem_+ \in \future$, the objective is to maximize the similarity between the user embedding $\embedding(\profile)$ and the positive item embedding $\embedding(\descript_{\ritem_+})$, while minimizing the similarity between $\embedding(\profile)$ and all other items in the batch (other items are considered as negative items):
\begin{align}
\embedding = \argmin_\embedding \sum_{\profile, B} - \ln \frac{\exp(sim(\embedding(\profile), \embedding(\descript_{\ritem_+})))}{\sum_{i \in B} \exp(sim(\embedding(\profile), \embedding(\descript_\ritem)))}
\label{eq:infonce}
\end{align}
By optimizing this contrastive loss, the model is trained to discriminate between relevant and irrelevant items, thereby improving the quality of recommendations generated for each user.

\subsubsection{Joint Optimization}

Combining the profile encoder optimization and recommender decoder optimization, we arrive at our joint optimization pipeline for \custompipelinename. The pseudocode is provided in Algorithm~\ref{alg:joint}. Our approach alternates between optimizing the two models over $K$ iterations. In each iteration, we first optimize the recommender decoder using a contrastive learning (\textbf{CL}) objective across $J$ batches, followed by optimizing the profile encoder via reinforcement learning for system optimization (\textbf{RLSO}) across $T$ batches. This alternating process ensures that both components are continuously refined, leading to a more cohesive and effective model for personalized recommendations. 

\section{Experiments}

In this section, we present a suite of experiments on various datasets and LLMs against multiple baselines to validate the effectiveness of \custompipelinename\ under different settings.

\subsection{Dataset}

We evaluate \custompipelinename\ on the Amazon-Movie-TV and Amazon-Books datasets~\citep{hou2024bridging} where the item metadata includes title, description, average rating, category, and price. In addition, we also include the user reviews of the items as additional inputs to the LLM. Examples of the item metadata and the LLM prompt are shown in Appendix~\ref{app:dataset_detail}. We filter out sessions and items that have less than 5 interactions. To satisfy the context length of the LLM and embedding model, we additionally filter out items with metadata that are longer than $512$ tokens and sessions with a history longer than $1024$ tokens.
The length of the interaction history is 4 and the number of items to predict in the future is 1 (i.e. $|h|=4$ and $|f|=1$), where it is guaranteed that all five items are distinct from each other.
The statistics of the dataset are shown in Table~\ref{tab:dataset_detail}.
To evaluate top-k ranking performance for each
model, we adopt three widely used metrics: Normalized Discounted Cumulative Gain (NDCG), Mean Reciprocal Rank (MRR), and Recall \citep{sun2019bert4recsequentialrecommendationbidirectional, gao2022mcl, lu2022sbr}.

{\renewcommand{\arraystretch}{1.0}
\begin{table}[t!]\centering
\resizebox{1.0\linewidth}{!}{
\begin{tabular}{c|ccc} 
\midrule[0.15ex]
\textbf{Dataset} & \makecell{\textbf{\#Sessions} \\ Train/Val/Test} & \textbf{\#Items} & \textbf{\#Interactions} \\ 
\midrule[0.05ex]
Amazon-Movie-TV & 14,698/1,392/1,392 & 10,533 & 87,410\\
Amazon-Books & 4,404/361/361 & 4,819 & 25,630\\
\midrule[0.15ex]
\end{tabular}}
\caption{Dataset statistics\label{tab:dataset_detail}}
\vspace{-2em}
\end{table}
}

\subsection{Baselines}

\subsubsection{Interpretable Profile Baselines}

We compare \custompipelinename\ with variants that reflect interpretable user profile methods from prior works. \textbf{\custompipelinename-0} conducts zero-shot inference using only pre-trained embedding and language models without additional training. This method represents the pure zero-shot approach described by \citet{zhou2024languagebaseduserprofilesrecommendation}. \textbf{\custompipelinename-CL} focuses on training the recommender decoder using InfoNCE, while keeping the profile encoder fixed. This baseline aligns with previous profile-based approaches in \citet{yang2023palrpersonalizationawarellms} and \citet{ ramos2024transparentscrutablerecommendationsusing}. To test the effectiveness of the profile encoder optimization, we include \textbf{\custompipelinename-RLSO} which only trains the encoder with RLSO without optimizing the decoder.

\subsubsection{Black-Box and No-Profile Baselines}

We also compare against a variety of other baseline types to contextualize the performance.

\textbf{No Profile methods} do not produce any profile for the user and serve as a baseline to assess task difficulty. These include \textbf{Random} which generates a randomly shuffled ranked list of all items, and \textbf{Most Popular (MP)} which ranks all items in descending order of popularity, based on observations from the training set and user interaction history from the validation/test set. These baselines serve as benchmarks for the task's difficulty, offering a point of reference for evaluating the performance of other methods.

\textbf{Black-Box Profile methods}\footnote{For all Black-Box Profile baselines, we use both the training set and the user interaction history items of the validation/test set during training.} rely on user and item interaction histories without leveraging additional item metadata. These methods apply collaborative filtering or other embedding-based approaches, producing user embeddings that are uninterpretable, hence referred to as black-box profiles.
\textbf{NMF}~\citep{NIPS2000_f9d11525} (Non-negative Matrix Factorization) decomposes user-item interaction data into latent factors and ranks the items based on the predictions of missing preferences.
\textbf{LightGCN}~\citep{he2020lightgcnsimplifyingpoweringgraph} uses Graph Convolutional Networks (GCN)~\citep{kipf2017semisupervisedclassificationgraphconvolutional} with Bayesian Personalized Ranking (BPR) loss~\citep{rendle2012bprbayesianpersonalizedranking} to learn robust user/item embeddings.
\textbf{MCL}~\citep{gao2022mcl} builds on top of LightGCN using sample mining and a mixed-centric loss.
\textbf{BERT4REC}~\citep{sun2019bert4recsequentialrecommendationbidirectional} uses masked language modeling (MLM) objective in BERT~\citep{devlin2019bertpretrainingdeepbidirectional} to learn user/item embeddings.
\textbf{LLM2BERT4REC}~\citep{Harte_2023} initializes then trains a BERT4REC model with embeddings obtained from a text embedding model using item descriptions.

\subsection{Implementation Details}

{\renewcommand{\arraystretch}{1.0}
\begin{table*}[t]\centering
\begin{tabular}[t]{cccccccc} 
\midrule[0.15ex]
\multirow{2}{*}{LLM} & \multirow{2}{*}{Method} & \multicolumn{3}{c}{Amazon-Movie-TV} & \multicolumn{3}{c}{Amazon-Books} \\
\cmidrule(lr){3-5} \cmidrule(lr){6-8}
& & Recall@20 & MRR@20 & NDCG@20 & Recall@20  &  MRR@20 & NDCG@20 \\
\midrule[0.05ex]
\parbox[t]{2mm}{\multirow{4}{*}{\rotatebox[origin=c]{90}{Gemma}}} & \custompipelinename-0 & 0.0643 &	0.0189 &	 0.0289 (+54.3\%*)  &	0.3397 & 0.1269 & 0.1743 (+57.0\%*) \\
& \custompipelinename-CL & \textbf{0.0863} &	\underline{0.0317} &  \underline{0.0436} (+2.29\% )& 0.3587  & 0.1516 & 0.1984 (+37.9\%*) \\
& \custompipelinename-RLSO & 0.0746 &	0.0253 &0.0364 (+22.5\%*)	  & \underline{0.4212} & \underline{0.1872} &  \underline{0.2400} (+14.0\%*)\\
& \custompipelinename & \underline{0.0859}  &	\textbf{0.0327}  &	\textbf{0.0446} & \textbf{0.4375} &	\textbf{0.2151} & \textbf{0.2736}	 \\
\midrule[0.05ex]
\parbox[t]{2mm}{\multirow{4}{*}{\rotatebox[origin=c]{90}{Llama}}} & \custompipelinename-0 & 0.0535  &	0.0146 &	0.0231 (+113.9\%*) & 0.3370  &	0.1285 &0.1753 (+36.5\%*)	 \\
& \custompipelinename-CL & \underline{0.0938} &	\underline{0.0319} &\underline{0.0456} (+8.33\%*)	  & \underline{0.3995} & 0.1539 &  0.2099 (+14.0\%*) \\
& \custompipelinename-RLSO & 0.0763 & 0.0289  &	0.0377 (+31.0\%*) & 0.3913 & \underline{0.1728} & \underline{0.2224} (+7.60\%*)  \\
& \custompipelinename & \textbf{0.0970} &	\textbf{0.0358}  &\textbf{0.0494}	 & \textbf{0.4212} & \textbf{0.1860} & \textbf{0.2393}	\\
\midrule[0.15ex]
\end{tabular} \\
\caption{\textbf{Results for Interpretable Profile Methods.} The best-performing methods for each metric, LLM and dataset are highlighted in bold and the next best methods are underlined. The relative improvements of \custompipelinename\ on NDCG@20 w.r.t. the baseline methods are shown in brackets. Asterisks denote statistically significant improvements based on paired t-test ($\alpha=0.05$). \label{tab:profile_result}}
\vspace{-1.5em}
\end{table*}}

We tune each baseline individually on the dataset to achieve the best performance on the validation set, and evaluate on the test set. We experiment with two different LLMs for the profile encoder, Gemma-2B-it\footnote{HuggingFace Model Card: google/gemma-2b-it}~\citep{gemmateam2024gemma2improvingopen} and Llama-3-8B-it\footnote{HuggingFace Model Card: meta-llama/Meta-Llama-3-8B-Instruct}~\citep{dubey2024llama3herdmodels}, to evaluate our framework across different LLM sizes and architectures. We set a maximum profile length of $512$ tokens. Following \citet{starling2023} and \citet{gao2024rebel}, we train the last four layers of the Llama-3-8B-it while keeping other layers frozen, and train Gemma-2B-it with Low-Rank Adaptors (LoRA)~\citep{hu2021loralowrankadaptationlarge}. Llama-3-8B-it experiments are trained on 8 A6000 GPUs for two days, and Gemma-2B-it experiments are trained on the same hardware for one day. For all methods that require a text embedding model, we use the same Mxbai\footnote{HuggingFace Model Card: mixedbread-ai/mxbai-embed-large-v1}~\citep{emb2024mxbai} embedding model, and perform full-parameter training on 8 A6000 GPUs with each iteration taking one hour.

In our implementation, we use NDCG as the ranking quality function $\util$ as it is bounded between 0 and 1. In addition, we standardize the values of $\util$ within each batch to zero-mean and unit-variance to further stabilize the learning of the policy. To ensure that $\encoder$ remains close to $\encoder_0$, we apply an additional KL penalty to the $\util$:
\begin{align}
&\util(\decoder(\profile, \mathcal{D}), \future) - \gamma\left(\ln \encoder(\profile|\metadata(\history)) - \ln \encoder_0(\profile|\metadata(\history))\right)
\end{align}
where $\gamma$ is a hyperparameter that controls the penalty~\citep{shengyi2022the37implementation, gao2024rebel}. Furthermore, to ensure that the generations terminate within the maximum generation length, following~\citet{shengyi2022the37implementation}, we penalize any generation that exceeds this length by setting $\util(\decoder(\profile, \mathcal{D}), \future)$ to a small fixed constant, $\Gamma$. Hyperparameter details of each method are shown in Appendix~\ref{app:hyper_details}.

\subsection{Performance Analysis}

\subsubsection{Performance Comparison among Interpretable Profile Methods}

The results for the interpretable profile methods are shown in Table~\ref{tab:profile_result} where the best-performing methods for each metric, base model and dataset are highlighted in bold and the second best methods are underlined. The relative improvements of \custompipelinename\ on NDCG@20 w.r.t the baselines are shown in bracket with asterisk denotes statistically significant improvement based on paired t-test ($\alpha=0.05$). While both \custompipelinename-CL and \custompipelinename-RLSO show substantial improvements over the non-learning approach \custompipelinename-0, \textbf{\custompipelinename\ achieves the highest performance compared to other interpretable profile methods in terms of NDCG and MRR}. This demonstrates \custompipelinename’s effectiveness: CL ensures user profiles and item metadata align in the embedding space, while RLSO refines the profile encoder, generating user profiles that better match target items, thereby improving recommendation accuracy. Examples of generated user profiles are shown in Figure~\ref{fig:example_data_and_profile} and Appendix~\ref{app:example_gen}, showing qualitatively that \custompipelinename\ is able to generate additional details compared to the base model while remaining interpretable.

{\renewcommand{\arraystretch}{1.0}
\begin{table*}[t]\centering
\begin{tabular}[t]{cccccccc} 
\midrule[0.15ex]
\multirow{2}{*}{Type} & \multirow{2}{*}{Method} & \multicolumn{3}{c}{Amazon-Movie-TV} & \multicolumn{3}{c}{Amazon-Books} \\
\cmidrule(lr){3-5} \cmidrule(lr){6-8}
& & Recall@20 & MRR@20 & NDCG@20 & Recall@20 & MRR@20 & NDCG@20 \\
\midrule[0.05ex]
\multirow{2}{*}{No Profile} & Random & 0.0019 & 0.0003 & 0.0006 (+8133\%*) & 0.0042 & 0.0007 & 0.0014 (+19443\%*)\\
& MP & 0.0129 & 0.0015 & 0.0039 (+1167\%*) & 0.0776 & 0.0714 & 0.0729 (+275.3\%*) \\
\midrule[0.05ex]
\multirow{5}{*}{Black-Box Profile} & NMF & 0.0625 &	0.0269 &	0.0350 (+41.1\%*) & {0.4238} & 0.1766 & 0.2323 (+17.8\%*) \\
& LightGCN & {0.1009} &	{0.0387} &	{0.0525} (-5.90\%*) & 0.4072 & {0.2145} & {0.2586} (+5.80\%*) \\
& MCL & {0.1078} &	{0.0417} &{0.0563} (-12.3\%*)	& 0.3906 & 0.1616 & 0.2134 (+28.2\%*) \\
& BERT4REC & 0.0758 & 0.0280 & 0.0384 (+28.6\%*) & 0.3006 & 0.1662 & 0.2235 (+22.4\%*) \\
& LLM2BERT4REC & 0.0783	& 0.0312 & 0.0423 (+16.8\%*) & 0.3572 & 0.2014 & 0.2343 (+16.8\%*) \\
\midrule[0.05ex]
Interpretable Profile & \custompipelinename & 0.0970 &	0.0358 & 0.0494 & {0.4375} &	{0.2151} &	{0.2736} \\
\midrule[0.15ex]
\end{tabular} \\
\caption{\textbf{Results using Different Recommender Types.} The relative improvements of \custompipelinename\ on NDCG@20 w.r.t. the baseline methods are shown in brackets. Asterisks denote statistically significant differences based on paired t-test ($\alpha=0.05$). \label{tab:result}}
\vspace{-1em}
\end{table*}}

\subsubsection{Performance Comparison with Black-Box Baselines}

We compare \custompipelinename\ with other state-of-the-art black-box recommendation methods in Table~\ref{tab:result}. 
For \custompipelinename, we select either the Llama or Gemma variant, depending on which yields the higher performance metric on the validation set.
Even though interpretability adds another objective and constraint, \custompipelinename\ outperforms all black-box methods on the Amazon-Books dataset, and only LightGCN and MCL manage to show better performance on Amazon-Movie-TV dataset. We attribute the excellent performance of \custompipelinename\ on Amazon-Books to the rich and informative textual descriptions in this dataset, which allows \custompipelinename\ to better capture user preferences. Additionally, it is possible that the LLMs have a deeper understanding of books compared to movies, resulting in more accurate profiles on Amazon-Books.

Overall, these results shows that \custompipelinename\ can be highly competitive even with methods that lack the interpretability and steerability of language-based profiles. These qualities are crucial for enhancing user experience and system control, making \custompipelinename\ more advantageous for applications that prioritize understanding and tailoring user preferences.


{\renewcommand{\arraystretch}{1.0}
\begin{table}[t]\centering
\resizebox{1.0\linewidth}{!}{\begin{tabular}[t]{ccccc} 
\midrule[0.15ex]
Method & Profile Length & Recall@20  & MRR@20& NDCG@20  \\
\midrule[0.05ex]
\multirow{4}{*}{\custompipelinename}  
& 64 & 0.0604  &	0.0201 &0.0319	 \\
& 128 &0.0901 &	0.0313 & 0.0444 	\\
& 256 & 	0.0929  &	0.0352 &0.0482 \\
& 512 &0.0970 & 0.0358 &0.0494	 \\
\midrule[0.05ex]
NoProfile-0 & N/A &0.0760  &	0.0261 &0.0372	 \\
NoProfile-CL & N/A &0.1012  &	0.0408 &0.0543		 \\
\midrule[0.15ex]
\end{tabular}}
\caption{\textbf{Ablation of Profile Length on Amazon-Movie-TV.}\label{tab:length_ab}}
\vspace{-1em}
\end{table}}

\subsection{Ablation Analysis}

\subsubsection{Profile Length}

The length of the profile plays an important role as it controls the width of the information bottleneck in the pipeline, where a shorter length requires stronger (and probably lossy) compression. We ablate the maximum generation length from 512 to 64 using Llama-3-8B-it on Amazon-Movie-TV. In addition, to quantify the amount of information loss through the bottleneck, we incorporate two additional baselines that have no bottleneck. \textbf{Zero-shot Embedding Model without Profile Encoder (NoProfile-0)} uses the same pre-trained Mxbai embedding model to generate user and item embeddings directly from the user interaction history with a list of item metadata and individual item metadata respectively. This approach is analogous to our pipeline illustrated in Fig.~\ref{fig:pipeline}, but without the profile encoder $\pi$ and the user profile $p$. \textbf{CL Embedding Model (NoProfile-CL)} is analogous to \textbf{NoProfile-0}, but the embedding model is trained with InfoNCE loss in Eq.~\ref{eq:infonce}.

The results are shown in Table~\ref{tab:length_ab}. We observe an almost monotonic increase in performance as the profile length increases, with the rate of improvement accelerating at shorter profile lengths. Profiles of length 256 and 512 show comparable performance, indicating that excessively long profiles are not necessary for optimal results. Beyond a certain length, increasing the profile size does not yield further improvements, as performance tends to plateau.

Compared to NoProfile-CL, we expected that the requirement to produce a language-based profile is a bottleneck that leads to marginally lower performance. NoProfile-CL avoids the bottleneck in our pipeline by directly utilizing the list of item metadata without incurring information loss. It is important to note that interpretable profile methods inherently experience some degree of information loss due to the generally shorter and more concise nature of profiles. Despite the slight reduction in performance, our approach offers significant advantages in terms of readability and interpretability, which we will demonstrate in Section~\ref{sec:interpretability}, as the generated profiles are far more understandable than a raw list of item metadata.

\subsubsection{Training Dataset Size}

\begin{figure}[t]
    \centering
    \includegraphics[scale=0.5,trim={5 5 5 5},clip]{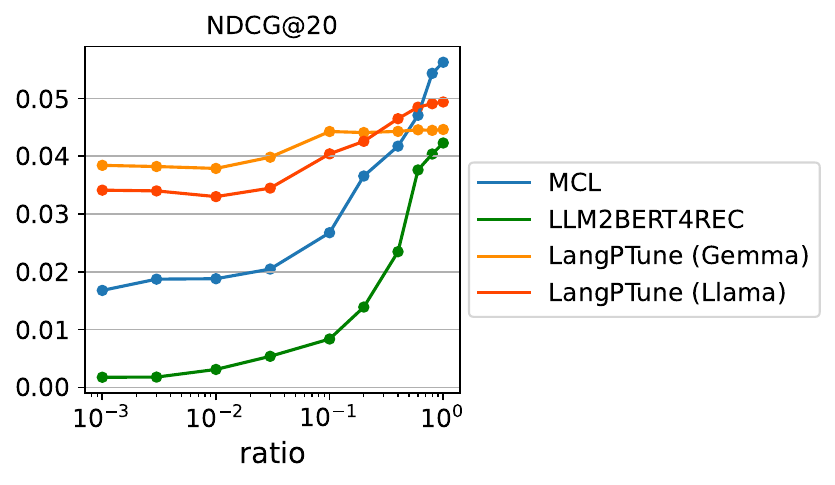}
    \vskip -0.3cm
    \caption{\textbf{Ablation of Train Data Ratio on Amazon-Movie-TV.} \label{fig:tran_ratio_ablation}}
    \Description[]{}
    \vspace{-2em}
\end{figure}

To assess the sensitivity of the methods to the amount of available training data, we conducted an ablation study by gradually reducing the training data from 100\% down to 0.1\% of the full training set size for Amazon-Movie-TV. The results are shown in Fig.~\ref{fig:tran_ratio_ablation}. MCL and LLM2BERT4REC experience significant performance degradation with reduced training data, as these methods require an explicit embedding for each item. In contrast, \custompipelinename\ is less affected, as it operates directly on text-based user profiles and item metadata. Their performance remains more consistent across the different dataset sizes. This suggests that interpretable profile methods may be particularly suitable for cold-start scenarios where there is little training data.

\begin{figure*}[t]
    \centering
    \includegraphics[scale=0.6,trim={60 150 40 160},clip]{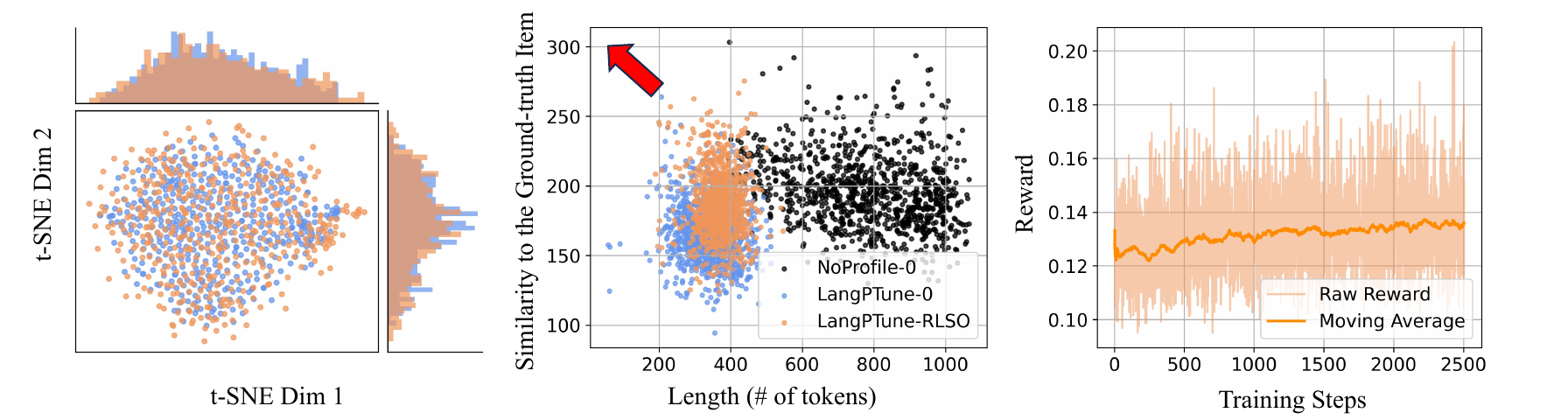}
    \vskip -0.3cm
    \caption{Visualizations with Llama-3-8B-it on Amazon-Movie-TV dataset. \textbf{(Left) Visualization of the embeddings of profiles from \custompipelinename-0 and \custompipelinename-RLSO using Mxbai and t-SNE. (Middle) Length vs. Similarity to the ground-truth item for the profiles from NoProfile-0, \custompipelinename-0, and \custompipelinename-RLSO. Red arrow indicates the optimal direction. (Right) Reward for \custompipelinename-RLSO during training.} \label{fig:visualizations}}
    \Description[]{}
    \vskip -0cm
\end{figure*}

\subsection{Interpretability Evaluation \label{sec:interpretability}}

Ideally, \custompipelinename\ should preserve the interpretability of the base model, ensuring that its generated profiles remain human-readable rather than generating a "secret code" that only the recommender decoder $\decoder$ can understand, as observed in the training of DeepSeek-R1-Zero~\citep{deepseekr1}.
To verify that \custompipelinename\ maintains interpretability of the generated profiles, we conduct a GPT-4 and human crowdworker study. 

\subsubsection{Dataset and Baselines}

We compare the profiles generated by \textbf{\custompipelinename}\ against two baselines: \textbf{\custompipelinename-0}, where profiles are generated through zero-shot inference using an LLM, and \textbf{NoProfile-0}, where the profile is a raw list of item metadata drawn from the user's interaction history. We utilize 450 items from the Amazon-Movie-TV test set and employ Llama-3-8B-it for \custompipelinename\ and \custompipelinename-0. Either a human or GPT-4 ({\texttt{gpt-4-0613}}) is presented with a profile text and asked to predict one of the two movies that the user would watch next. The two-movie pool for each user consists of the ground truth movie randomly paired with another movie from the test set. The negative movie option for each user remains consistent across profile types. This experiment setup allows us to quantitatively assess the interpretability and usefulness of the generated profiles, since better profiles should lead to more accurate predictions of the next movie by humans and general-purpose language models (e.g. GPT-4) other than the recommender decoder $\decoder$. Further detail is available in Appendix \ref{app:survey_details}.


{\renewcommand{\arraystretch}{1.0}
\begin{table}[t]\centering
\resizebox{0.55\linewidth}{!}{\begin{tabular}[t]{ccc} 
\midrule[0.15ex]
Profile Type & Human & GPT-4  \\
\midrule[0.05ex]
NoProfile-0 & 57.11\% & 75.56\% \\
\custompipelinename-0 & 54.67\% & 73.33\% \\
\custompipelinename & \textbf{58.89\%} & \textbf{76.67\%} \\
\midrule[0.15ex]
\end{tabular}}
\caption{\textbf{Interpretability task average accuracy rates with human participants and GPT-4 responses. Highest accuracy per respondent type highlighted in bold.}\label{tab:interpret_analysis}}
\vspace{-2em}
\end{table}}

\subsubsection{Results}

We compare human and GPT-4 accuracy rates across the three different profile types in Table~\ref{tab:interpret_analysis}. We find that both human participants and GPT-4 achieve higher accuracy on this inference task with \custompipelinename\ profiles than with \custompipelinename-0 profiles. Although the difference between \custompipelinename\ and \custompipelinename-0 on human survey responses is not significant via two-sided McNemar's test with exact Bernoulli p-value ($p=0.201$), the ordinal alignment of the human and GPT-4 results suggests that the profiles generated by \custompipelinename\ are at least as interpretable as those generated by \custompipelinename-0.


We also compare the average profile lengths and response durations for human participants in Table~\ref{tab:interpret_analysis_duration}. Although the performance difference between \custompipelinename\ and NoProfile-0 on human surveys is smaller ($p=0.624$), NoProfile-0 has much longer texts and requires more time to complete, indicating that its raw format imposes a higher cognitive load on participants. In contrast, \custompipelinename\ profiles are more concise while preserving the essential information needed for accurate inference.

\subsection{Additional Analysis}

To gain deeper insights into our pipeline, we present three visualizations based on experiments with Llama-3-8B-it using the Amazon-Movie-TV dataset, as shown in Fig.~\ref{fig:visualizations}. These visualizations focus on \custompipelinename-0 and \custompipelinename-RLSO, as neither involves training the embedding model. This allows us to use the same zero-shot Mxbai embedding model to extract their profile embeddings, ensuring a consistent basis for comparison. We first visualize the profile embeddings generated by \custompipelinename-0 and \custompipelinename-RLSO, with t-SNE\cite{Maaten2008VisualizingDU} for dimensionality reduction. The results indicate that \custompipelinename-RLSO produces more detailed profiles, as reflected in its broader distribution. We also compare the number of tokens in profiles from NoProfile-0 (i.e. a raw list of item metadata), \custompipelinename-0, and \custompipelinename-RLSO against their similarity to ground-truth items. Similarity is measured via the dot product of embeddings generated by Mxbai between the profile and the item. While \custompipelinename-RLSO slightly increases profile length, it significantly enhances similarity to the ground-truth items, reaching levels comparable to NoProfile-0. Finally, we examine the reward trajectory during \custompipelinename-RLSO training. The results show a steady and consistent enhancement in recommendation performance without any performance rollbacks.

{\renewcommand{\arraystretch}{1.0}
\begin{table}[t]\centering
\resizebox{1.0\linewidth}{!}{\begin{tabular}[t]{cccc} 
\midrule[0.15ex]
Profile/Text Type & Avg Char \# & Avg Word \# & Avg Duration (sec)  \\
\midrule[0.05ex]
(Answer Options) & 701 & 108 & - \\
NoProfile-0 & 2983 & 465 & 34.76 \\
\custompipelinename-0 & 1598 & 292 & 28.77 \\
\custompipelinename & 1730 & 256 & 29.15 \\
\midrule[0.15ex]
\end{tabular}}
\caption{\textbf{Interpretability task duration for human participants with average text lengths ($|z|>3$ outliers filtered out)}\label{tab:interpret_analysis_duration}}
\vspace{-1em}
\end{table}}

\section{Conclusions and Future Work}

We introduce \custompipelinename, the first end-to-end pipeline that optimizes LLMs to generate user profiles for recommendation. By combining Reinforcement Learning for System Optimization (RLSO) with contrastive learning (CL), our approach outperforms prior language-based profile methods and matches black-box recommendation models while maintaining interpretability. Human and GPT-4 evaluations confirm that \custompipelinename\ generates readable and informative profiles. Furthermore, there is a number of immediate research questions. First, while we evaluated interpretability, it would also be interesting to design user studies that evaluate steerability through manual changes to the user profiles. Second, on the technical side, instead of only focusing on user profile generation with clear context lengths, it would be interesting to adapt our pipeline to accommodate longer interaction histories or to obtain better natural-language item descriptions.

\newpage
\section{Acknowledgments}
This research was supported in part by the Graduate Fellowships for STEM Diversity (GFSD) and LinkedIn-Cornell Grant, as well as NSF Awards IIS-2008139 and IIS-2312865.
All content represents the opinion of the authors, which is not necessarily shared or endorsed by their respective employers and/or sponsors.

\bibliographystyle{ACM-Reference-Format}
\bibliography{reference}

\newpage
\appendix
\onecolumn
\section{Dataset Details}
\label{app:dataset_detail}

The session description consists of a sequence of item metadatas in the session. Examples of a session description and an item metadata are shown below:

\begin{table*}[htb]\centering
\raggedright{\textbf{Amazon-Movie-TV $|$ Session Description Example}}
\resizebox{\linewidth}{!}{
\begin{tabular}{p{1.1\linewidth}} 
\midrule[0.3ex]
\textbf{Title}: Top Gun [DVD \& Digital Copy Included] [3D Blu-ray] \newline
\textbf{Description}: (average rating: 4.7) Top Gun takes a look at the danger and excitement that awaits every pilot at the Navy's prestigious fighter weapons school. Tom Cruise is superb as Pete "Maverick" Mitchell, a daring young flyer who's out to become the best of the best. And Kelly McGillis sizzles as the civilian instructor who teaches Maverick a few things you can't learn in a classroom. \newline
\textbf{Category}: Movies \& TV, Blu-ray, Movies \newline
\textbf{Price}: 17.4 \newline
\textbf{Review from the user}: (rating: 1.0) Always liked this movie and was excited to see that it was be re-released in 3D BluRay.  Well it is 3D, but the quality is horrible.  The picture in some spots looks like VHS.  The picture in the best spots looks no better than DVD.<br /><br />I'm learning that I shouldn't spend good money to upgrade DVD films to 3D or Blu Ray. \newline
 \newline
\textbf{Title}: Person of Interest: Season 4 (Blu-ray + Digital Copy) \newline
\textbf{Description}: (average rating: 4.8) Person of Interest: The Complete Fourth Season (Blu-ray) ]]> \newline
\textbf{Category}: Movies \& TV, Blu-ray, TV \newline
\textbf{Price}: 53.66 \newline
\textbf{Review from the user}: (rating: 4.0) Good, but not as good as the first three seasons.  Too much about AI battles and not enough about people. \newline
 \newline
\textbf{Title}: The Martian \newline
\textbf{Description}: (average rating: 4.8) From legendary director Ridley Scott (Alien, Prometheus) comes a gripping tale of human strength and the will to survive, starring Matt Damon as an astronaut stranded on Mars. \newline
\textbf{Category}: Science Fiction, Drama, Anxious, Biting \newline
\textbf{Price}: 4.99 \newline
\textbf{Review from the user}: (rating: 5.0) A really good movie.  Well done in all aspects.  Great story, great acting, great visual effects.  Thoroughly enjoyed it. \newline
 \newline
\textbf{Title}: The Firm: The Complete Series \newline
\textbf{Description}: (average rating: 4.3) As a young associate Mitch brought down Bendini Lambert \& Locke a prestigious Memphis law firm operating as a front for the Chicago Mob. His life was never the same. After a difficult ten years which included a stay in the Federal Witness Protection Program Mitch and his family emerge from isolation determined to reclaim their lives and their future -- only to find that past dangers are still lurking and new threats are everywhere.NOTE:Disc 1 is behind the insert attached to back of front cover. \newline
\textbf{Category}: Movies \& TV, Boxed Sets, Drama \newline
\textbf{Price}: None \newline
\textbf{Review from the user}: (rating: 4.0) Loved most of it.  Got a little predictable towards the end.  Great acting.
\\
\midrule[0.3ex]
\end{tabular}}
\end{table*}

\begin{table*}[htb]\centering
\raggedright{\textbf{Amazon-Movie-TV $|$ Item Metadata Example}}
\resizebox{\linewidth}{!}{
\begin{tabular}{p{1.1\linewidth}} 
\midrule[0.3ex]
\textbf{Title}: Outlander: Season 5 [DVD] \newline
\textbf{Description}: (average rating: 4.9) The fifth season of Outlander sees a continuation of Claire and Jamie’s fight to protect those they love, as they navigate the trials and tribulations of life in colonial America. The Frasers strive to flourish within a society which is unwittingly marching toward Revolution, and Jamie must now defend the home they have built together at Fraser’s Ridge while Claire seeks to put her own skills and medical expertise to use in keeping her family together and safe from harm. Meanwhile, Brianna and Roger MacKenzie struggle to find their respective places in this world and chase away the shadow cast over their lives by Stephen Bonnet as they raise their son in this brave new world. For the Frasers and their family, “home” is more than simply a site on which they live, it is the place in which they are laying the foundations for the rest of their lives. \newline
\textbf{Category}: Movies \& TV, Studio Specials, Sony Pictures Home Entertainment, All Sony Pictures Titles \newline
\textbf{Price}: 17.61 \\
\midrule[0.3ex]
\end{tabular}}
\end{table*}

\begin{table*}[htb]\centering
\raggedright{\textbf{Amazon-Movie-TV $|$ Prompt for Generation}}
\resizebox{\linewidth}{!}{
\begin{tabular}{p{1.1\linewidth}} 
\midrule[0.3ex]
The user has rated (out of 5.0) and reviewed following movies and TV shows arranged chronologically from the oldest (top) to the newest (bottom). Please provide a high-level summary of the user preference in detail. \\
\midrule[0.3ex]
\end{tabular}}
\end{table*}

\newpage

\begin{table*}[htb]\centering
\raggedright{\textbf{Amazon-Books $|$ Session Description Example}}
\resizebox{\linewidth}{!}{
\begin{tabular}{p{1.1\linewidth}} 
\midrule[0.3ex]
\textbf{Title}: Delvers LLC: Adventure Capital \newline
\textbf{Description}: (average rating: 4.6) From the Author Delvers LLC: Adventure Capital is quite possibly the most heroic, the darkest, the most action-packed, and the most min/maxing installment of Delvers yet!As most know, Delvers LLC is a series of portal fiction, where regular people are transported to a fantasy world.  However, where other series hand wave away a lot of real world issues, Delvers LLC explores them.  Turning tropes sideways is a perverse hobby of mine. :)I am very confident that readers are going to love this book.  Either way, I wish I could thank each and every reader individually for coming along with Henry and Jason on this journey. \newline
\textbf{Category}: Books, Literature \& Fiction, Genre Fiction \newline
\textbf{Price}: 0.0 \newline
\textbf{Review from the user}: (rating: 5.0) This story moves at a fast pace and is definitely engaging and enjoyable. That said the ending of the book did not actually keep up the same type of tension as book two. Will definitely still pick up book 4 \newline\newline
\textbf{Title}: Metal Mage 2 (Metal Mage (Completed Series)) \newline
\textbf{Description}: (average rating: 4.5) \newline
\textbf{Category}: Books, Literature \& Fiction, Genre Fiction \newline
\textbf{Price}: 0.0 \newline
\textbf{Review from the user}: (rating: 5.0) Like others have said this book is a little slower than the first one. Still the story is overall good. I would say that it doesn't really shock you, and there never really comes a time when you doubt that Mason will be victorious, which is a shame, but it's good for what it is \newline\newline
\textbf{Title}: Metal Mage 4 (Metal Mage (Completed Series)) \newline
\textbf{Description}: (average rating: 4.6) \newline
\textbf{Category}: Books, Literature \& Fiction, Genre Fiction \newline
\textbf{Price}: 0.0 \newline
\textbf{Review from the user}: (rating: 4.0) The story here is not as fast paced or interesting as before. I would almost call its formulaic in parts. Mason finally makes it to the dwarves and of course things don't go swimmingly at first, but he inevitably wins them over. Really once again he never comes anywhere close to losing, which is getting old.<br />Still the book is an enjoyable read minus one REALLY annoying pattern the author is getting into.<br /><br />I am aware that this series idea not primarily about sex or spelling out the sex scenes he creates. However the author is REALLY bad at starting things that seem really interesting, and then simply fading to black after a page. This is really annoying. Either go the distance or leave it out of the book, because otherwise it's just pointless and distracting. \newline\newline
\textbf{Title}: Demon Driven (The Demon Accords Book 2) \newline
\textbf{Description}: (average rating: 4.6) About the Author John Conroe was raised in Saint Lawrence County, New York, a few heartbeats away from the Canadian border at the very top of the state. It is a wild and rural place where most rivers run south to north and the inhabitants all sound vaguely like their national flag should feature a massive red maple leaf instead of stars and stripes. James Patrick Cronin is an accomplished audiobook narrator with over 100 titles to his credit. A classically trained stage actor with an MFA from the University of Louisville, he has performed as an actor and a comedian on stages all over the world. --This text refers to an out of print or unavailable edition of this title. \newline
\textbf{Category}: Books, Literature \& Fiction, Genre Fiction \newline
\textbf{Price}: 0.0 \newline
\textbf{Review from the user}: (rating: 3.0) This book isn't bad. The story is interesting enough, and the journey isn't terrible. The issue start to become apparent pretty early however. In the first book the main character has some noticeable challenges, but this in this book they are no where to be seen. I like to think about this as similar to why stories about Superman are hard to do. The MC isn't quite at Superman levels yet, but he's in his way there. There are several problems put in his way, but they are all solved with ease, which gets boring. The only time in this book that gets truly tense is when something happens to Tanya, but again it's easily fixed and things move on quickly.<br />Look, I'm a little short on things to read right now so I'm going to continue this series. However if the books continue to read like amateur fan fiction I just won't be able to keep going.
\\
\midrule[0.3ex]
\end{tabular}}
\vskip -0.3cm
\end{table*}

\begin{table*}[htb]\centering
\raggedright{\textbf{Amazon-Books $|$ Item Metadata Example}}
\resizebox{\linewidth}{!}{
\begin{tabular}{p{1.1\linewidth}} 
\midrule[0.3ex]
\textbf{Title}: Tokyo Ghoul, Vol. 4 (4) \newline
\textbf{Description}: (average rating: 4.8) About the Author Sui Ishida was born in Fukuoka, Japan. He is the author of Tokyo Ghoul and several Tokyo Ghoul one-shots, including one that won him second place in the Weekly Young Jump 113th Grand Prix award in 2010. Tokyo Ghoul began serialization in Weekly Young Jump in 2011 and was adapted into an anime series in 2014. \newline
\textbf{Category}: Books, Comics \& Graphic Novels, Manga \newline
\textbf{Price}: 4.69 \\
\midrule[0.3ex]
\end{tabular}}
\vskip -0.3cm
\end{table*}

\begin{table*}[htb]\centering
\raggedright{\textbf{Amazon-Books $|$ Prompt for Generation}}
\resizebox{\linewidth}{!}{
\begin{tabular}{p{1.1\linewidth}} 
\midrule[0.3ex]
The user has rated (out of 5.0) and reviewed following books arranged chronologically from the oldest (top) to the newest (bottom). Please provide a high-level summary of the user preference in detail. \\
\midrule[0.3ex]
\end{tabular}}
\end{table*}

\newpage

\section{Hyperparameter Details}
\label{app:hyper_details}

The hyperparameters are tuned on validation dataset.

\begin{table*}[htb!]\centering
\resizebox{\linewidth}{!}{
\begin{tabular}{p{0.3\linewidth}p{0.3\linewidth}p{0.3\linewidth}}
\midrule[0.3ex]
\textbf{Method} &
\textbf{Parameters} \\
\midrule[0.15ex]
Non-negative matrix factorization (NMF) & 
Components: 1000 \newline
Solver: batch HALS &
Beta loss: L2 \\
\midrule[0.15ex]
LightGCN & 
Learning rate: 1e-3 \newline
Batch size: 1000 \newline
Dimension: 512 &
Layer: 3 \newline
Num negative: 1 \\
\midrule[0.15ex]
MCL & 
Learning rate: 1e-3 \newline
Batch size: 1000 \newline
Dimension: 512 \newline
Layer: 3 \newline
Num negative: 10 &
$\alpha$: 1.25 \newline
$\beta$: 5.0 \newline
$\lambda_p$: 6.5 \newline
$\lambda_n$: -0.5 \\
\midrule[0.15ex]
BERT4REC & 
Batch size: 128 \newline
Learning rate: 1e-3 \newline
Decay step: 25 \newline
Epochs: 100 \newline
Dropout: 0.1 & 
Hidden units: 512 \newline
Mask prob: 0.5 \newline
Num blocks: 2 \newline
Num heads: 4 \\
\midrule[0.15ex]
LLM2BERT4REC &
Batch size: 128 \newline
Learning rate: 1e-3 \newline
Decay step: 25 \newline
Epochs: 100 \newline
Dropout: 0.1 & 
Hidden units: 512 \newline
Mask prob: 0.5 \newline
Num blocks: 2 \newline
Num heads: 4 \\ 
\midrule[0.15ex]
NoProfile-CL & 
CL learning rate: 1e-5 \newline
CL batch size: 256 \newline
CL weight decay: 1e-6 \newline
Schedule: linear decay &
$K$: 5 \newline
$J$: 800 \newline
Embedding size: 512 \\
\midrule[0.15ex]
\custompipelinename-CL & 
CL learning rate: 1e-5 \newline
CL batch size: 256 \newline
CL weight decay: 1e-6 \newline
Schedule: linear decay &
$K$: 5 \newline
$J$: 800 \newline
Embedding size: 512 \\
\midrule[0.15ex]
\end{tabular}}
\end{table*}

\newpage
\begin{table*}[htb!]\centering
\resizebox{\linewidth}{!}{
\begin{tabular}{p{0.3\linewidth}p{0.3\linewidth}p{0.3\linewidth}}
\midrule[0.15ex]
\custompipelinename-RLSO (Gemma-2B-it) & 
RLSO learning rate: 1e-7 \newline
RLSO batch size: 64 \newline
RLSO weight decay: 1e-6 \newline
Schedule: linear decay \newline 
LoRA rank: 1024 \newline
LoRA alpha: 2048 \newline
$T$: 200 &
$K$: 5 \newline
Num epochs: 4 \newline
$\eta$: 1.0 \newline
$\gamma$: 0.2 \newline
$\Gamma$: 0 \newline 
Embedding size: 512 \\
\midrule[0.15ex]
\custompipelinename-RLSO (Llama-3-8B-it) & 
RLSO learning rate: 3e-7 \newline
RLSO batch size: 64 \newline
RLSO weight decay: 1e-6 \newline
Schedule: linear decay \newline
$T$: 500 \newline
$K$: 5 &
Num epochs: 4 \newline
$\eta$: 1.0 \newline
$\gamma$: 0 \newline
$\Gamma$: 0 \newline
Embedding size: 512 \\
\midrule[0.15ex]
\custompipelinename\ (Gemma-2B-it) & 
RLSO learning rate: 1e-7 \newline
RLSO batch size: 64 \newline
RLSO weight decay: 1e-6 \newline
CL learning rate: 1e-7 \newline
CL batch size: 256 \newline
CL weight decay: 1e-6 \newline
Schedule: linear decay \newline 
LoRA rank: 1024 \newline
LoRA alpha: 2048 &
$T$: 200 \newline
$K$: 5 \newline
$J$: 800 \newline
Num epochs: 4 \newline
$\eta$: 1.0 \newline
$\gamma$: 0.2 \newline
$\Gamma$: 0 \newline 
Embedding size: 512 \\
\midrule[0.15ex]
\custompipelinename\ (Llama-3-8B-it) & 
RLSO learning rate: 3e-7 \newline
RLSO batch size: 64 \newline
RLSO weight decay: 1e-6 \newline
CL learning rate: 3e-7 \newline
CL batch size: 256 \newline
CL weight decay: 1e-6 \newline
Schedule: linear decay \newline
$T$: 500 &
$K$: 5 \newline
$J$: 2500 \newline
Num epochs: 4 \newline
$\eta$: 1.0 \newline
$\gamma$: 0 \newline
$\Gamma$: 0 \newline 
Embedding size: 512\\
\midrule[0.3ex]
\end{tabular}}
\end{table*}

\clearpage
\section{Example Generations}
\label{app:example_gen}

In this section, we show different model generations given the following input prompt:

\begin{table*}[htb]\centering
\raggedright{\textbf{Input Prompt}}
\resizebox{\linewidth}{!}{
\begin{tabular}{p{1.1\linewidth}} 
\midrule[0.3ex]
The user has rated (out of 5.0) and reviewed following movies and TV shows arranged chronologically from the oldest (top) to the newest (bottom). Please provide a high-level summary of the user preference in detail.\newline
\newline
\textbf{Title}: Life of Pi (Blu-ray + DVD + Digital Copy)\newline
\textbf{Description}: (average rating: 4.7) Director Ang Lee creates a groundbreaking movie event about a young man who survives a tragic disaster at sea and is hurtled into an epic journey of adventure and discovery. While marooned on a lifeboat, he forms an amazing and unexpected connection with the ship's only other survivor...a fearsome Bengal tiger.\newline
\textbf{Category}: Movies \& TV, Featured Categories, Blu-ray, Action \& Adventure\newline
\textbf{Price}: 7.99\newline
\textbf{Review from the user}: (rating: 3.0) great story, great visuals and great message.  It's a long movie and some parts might be boring but good movie overall.  Also there are more subtitles than indicated... I think it's subtitled with every language in the world!\newline
\newline
\textbf{Title}: 10 Cloverfield Lane\newline
\textbf{Description}: (average rating: 4.7) A young woman wakes up after a terrible accident to find that she's locked in a cellar with a doomsday prepper, who insists that he saved her life and that the world outside is uninhabitable following an apocalyptic catastrophe. Uncertain what to believe, the woman soon determines that she must escape at any cost. Movie stars John Goodman, Mary Elizabeth Winstead, John Gallagher Jr and Bradley Cooper.\newline
\textbf{Category}: Movies \& TV, Featured Categories, Blu-ray, Horror
\textbf{Price}: 9.99\newline
\textbf{Review from the user}: (rating: 5.0) great movie, very suspenseful and claustrophobic.  kept me guessing to the end.  never seen the original Cloverfield so didn't know what to expect, so of course the ending was a bit unexpected.  acting is superb.\newline
\newline
\textbf{Title}: Inception (Rpkg/BD) [Blu-ray]\newline
\textbf{Description}: (average rating: 4.6) Inception (Rpkg/BD) Acclaimed filmmaker Christopher Nolan directs an international cast in this sci-fi actioner that travels around the globe and into the world of dreams. Dom Cobb (Leonardo DiCaprio) is the best there is at extraction: stealing valuable secrets inside the subconscious during the mind’s vulnerable dream state. His skill has made him a coveted player in industrial espionage but also has made him a fugitive and cost him dearly. Now he may get a second chance if he can do the impossible: inception, planting an idea rather than stealing one. If they succeed, Cobb and his team could pull off the perfect crime. But no planning or expertise can prepare them for a dangerous enemy that seems to predict their every move. An enemy only Cobb could have seen coming. ]]>\newline
\textbf{Category}: Movies \& TV, Blu-ray, Movies\newline
\textbf{Price}: 12.88\newline
\textbf{Review from the user}: (rating: 5.0) it's an awesome movie and surely most people know that already\newline
\newline
\textbf{Title}: Kingsman: The Secret Service (Blu-ray + Digital Copy)\newline
\textbf{Description}: (average rating: 4.7) A Super-Secret Organization recruits an unrefined but promising street kid into the agency's ultra-competitive training program just as a dire global threat emerges from a twisted tech genius. A phenomenal cast, including Academy Award Winners Colin Firth and Michael Caine and Oscar Nominee Samuel L. Jackson, leads this action-packed spy-thriller directed by Matthew Vaughn (X-Men: First Class).\newline
\textbf{Category}: Movies \& TV, Blu-ray, Movies\newline
\textbf{Price}: 8.99\newline
\textbf{Review from the user}: (rating: 5.0) Epic action film, recommended for action film fans.\\
\midrule[0.3ex]
\end{tabular}}
\end{table*}

\begin{table*}[htb]\centering
\raggedright{\textbf{Gemma-2B-it} $|$ $k=5$, Maximum Generation Length = 512}
\resizebox{\linewidth}{!}{
\begin{tabular}{p{1.1\linewidth}} 
\midrule[0.3ex]
User Preference Summary:\newline
\textbf{Life of Pi (Blu-ray + DVD + Digital Copy)}: The user finds the movie enjoyable with visually captivating visuals, a thought-provoking story, and a positive message about courage and resilience. However, the lengthy runtime may be challenging for some viewers.\newline
\textbf{10 Cloverfield Lane}: The user appreciates the suspenseful and claustrophobic atmosphere, as well as the unexpected ending that reveals the reality behind the apocalypse. The movie is a unique and refreshing take on a dystopian world.\newline
\textbf{Inception (Rpkg/BD)}: The detailed and visual nature of the dream-world sequences in Inception is highly appreciated by the user. The film features exceptional acting and a thought-provoking narrative that explores the complexities of dreams and reality.\newline
\textbf{Kingsman}: The Secret Service (Blu-ray + Digital Copy):** The user finds the movie thrilling and fast-paced, with an exhilarating action film combined with humorous comedic elements. The cast of skilled actors and the movie's fast-paced editing contribute to the overall enjoyment.\newline
\textbf{Overall, the user has a diverse and positive taste in movies and TV shows, with a particular fondness for visual storytelling and thought-provoking narratives.} \\
\midrule[0.3ex]
\end{tabular}}
\end{table*}

\begin{table*}[htb]\centering
\raggedright{\textbf{Llama-3-8B-it} $|$ $k=5$, Maximum Generation Length = 512}
\resizebox{\linewidth}{!}{
\begin{tabular}{p{1.1\linewidth}} 
\midrule[0.3ex]
Based on the user's ratings and reviews, here is a summary of their preferences:\newline
\textbf{Storytelling}: The user enjoys complex storylines with twists and surprises. They appreciate movies that explore themes of adventure, discovery, and suspense. The user particularly liked the unexpected connections and surprises in ""Life of Pi"" and the twists in ""10 Cloverfield Lane"".\newline
\textbf{Visuals}: The user appreciates visually stunning movies with groundbreaking effects. They praised the visuals in ""Life of Pi"" and might have appreciated the action sequences and spy-thriller elements in ""Kingsman: The Secret Service"".\newline
\textbf{Suspense and Thrills}: The user enjoys suspenseful movies that keep them guessing until the end. They highly rated ""10 Cloverfield Lane"" for its ability to maintain tension and uncertainty throughout the film.\newline
\textbf{Acting:} The user values strong acting performances, praising the talents of the cast in ""10 Cloverfield Lane"" and potentially appreciating the acting in ""Kingsman: The Secret Service"".\newline
\textbf{Blu-ray and Extras}: The user is willing to pay a premium price for Blu-ray copies of movies, suggesting that they appreciate high-quality visual and sound presentation.\newline
\textbf{Movie Length}: The user might be willing to tolerate longer movie runtime if the story is engaging, as evidenced by their 3.0 rating for ""Life of Pi"", which they found was ""a long movie"".\newline
\textbf{Subtitles}: The user might be sensitive to subtitles, noting that there were more subtitles than indicated in ""Life of Pi"".\newline
\textbf{Action and Adventure}: The user enjoys action-packed movies, particularly spy-thrillers like ""Kingsman: The Secret Service"" and adventure films like ""10 Cloverfield Lane"".\newline
\textbf{Sci-Fi and Fantasy}: The user might appreciate sci-fi elements, as evidenced by their appreciation for the themes of dreams and reality in ""Inception"".\newline
\textbf{Classics}: The user might be willing to watch classics, but they didn't mention any particular appreciation for them.\newline
Overall, the user enjoys complex storylines, suspenseful plots, and strong acting performances, particularly in action-packed and sci-fi thrillers.\\
\midrule[0.3ex]
\end{tabular}}
\end{table*}

\begin{table*}[htb]\centering
\raggedright{\textbf{Llama-3-8B-it} $|$ $k=5$, Maximum Generation Length = 256}
\resizebox{\linewidth}{!}{
\begin{tabular}{p{1.1\linewidth}} 
\midrule[0.3ex]
Based on the ratings and reviews provided, it appears that the user prefers suspenseful and action-packed movies with engaging storylines and strong suspense elements. They have rated:\newline
Life of Pi (Blu-ray + DVD + Digital Copy) - 3/5 stars\newline
10 Cloverfield Lane - 5/5 stars\newline
Inception (Rpkg/BD) - 5/5 stars\newline
Kingsman: The Secret Service (Blu-ray + Digital Copy) - 5/5 stars\newline
The user praised the suspensefulness and claustrophobic atmosphere of 10 Cloverfield Lane, the action-packed nature of Kingsman: The Secret Service, and the engaging storytelling and suspense elements of Inception. They rated Life of Pi relatively highly, praising its story, visuals, and message, but criticized its length and the abundance of subtitles.\newline
Overall, the user prefers movies with strong suspense elements, engaging storylines, and fast-paced action sequences.\\
\midrule[0.3ex]
\end{tabular}}
\end{table*}

\begin{table*}[htb]\centering
\raggedright{\textbf{Llama-3-8B-it} $|$ $k=5$, Maximum Generation Length = 128}
\resizebox{\linewidth}{!}{
\begin{tabular}{p{1.1\linewidth}} 
\midrule[0.3ex]
Based on the ratings and reviews, it is clear that the user has a strong preference for action-based Blu-ray movies, Blu-ray + Digital Copy films, and Blu-ray movies, such as ""Life of Pi"", ""10 Cloverfield Lane"", ""Inception"", ""Kingsman: The Secret Service"".\\
\midrule[0.3ex]
\end{tabular}}
\end{table*}

\begin{table*}[htb]\centering
\raggedright{\textbf{Llama-3-8B-it} $|$ $k=5$, Maximum Generation Length = 64}
\resizebox{\linewidth}{!}{
\begin{tabular}{p{1.1\linewidth}} 
\midrule[0.3ex]
Based on the ratings and reviews provided, it appears that the user has a preference for action-adventure movies, sci-fi, and suspenseful horror films.\\
\midrule[0.3ex]
\end{tabular}}
\end{table*}

\begin{table*}[htb]\centering
\raggedright{\textbf{Llama-3-8B-it} $|$ $k=4$, Maximum Generation Length = 512}
\resizebox{\linewidth}{!}{
\begin{tabular}{p{1.1\linewidth}} 
\midrule[0.3ex]
Based on the user's ratings and reviews, here is a summary of their preferences : \newline
\textbf{Storytelling}: the user appreciates complex and suspenseful storylines, as evidenced by their ratings of 5. 0 for 10 cloverfield lane and kingsman : the secret service, and 3. 0 for life of pi. they enjoy plots that keep them guessing until the end. \newline
\textbf{Action and Suspense}: the user is drawn to action - packed films, as demonstrated by their ratings of 5. 0 for kingsman : the secret service, 10 cloverfield lane, and inception. they seem to enjoy fast - paced plots with twists and turns. \newline
\textbf{Visuals}: the user appreciates visually stunning films, as indicated by their ratings of 3. 0 for life of pi, which they praised for its great visuals. \newline
\textbf{Acting}: the user values good acting, as evidenced by their ratings of 5. 0 for 10 cloverfield lane, which they praised for superb acting. \newline
\textbf{thematic complexity}: the user may be drawn to films that deal with complex themes, as suggested by their appreciation for inception's exploration of reality and perception. \newline
\textbf{Blu-ray Format}: the user appears to prioritize blu - ray formats, as they purchased and watched all the blu - ray movies on the list. \newline
\textbf{Budget-friendly Options}: the user may be budget - conscious, as they seem to be willing to pay reasonable prices for blu-ray copies of movies, ranging from 7.99 to 12.88. \newline
Overall, the user enjoys complex storylines, action - packed plots, suspenseful twists, and visually stunning films, but may be willing to overlook some pacing issues if the story and themes are compelling.\\
\midrule[0.3ex]
\end{tabular}}
\end{table*}

\begin{table*}[htb]\centering
\raggedright{\textbf{Llama-3-8B-it} $|$ $k=3$, Maximum Generation Length = 512}
\resizebox{\linewidth}{!}{
\begin{tabular}{p{1.1\linewidth}} 
\midrule[0.3ex]
Based on the user's ratings and reviews, here is a summary of their preferences : \newline
\textbf{movie genre}: the user enjoys action - packed movies with a mix of adventure, suspense, and thriller elements. they appreciate movies that keep them guessing until the end and have a strong plot. \newline
\textbf{storytelling}: the user values storytelling over special effects, as evident from their appreciation for movies like "" life of pi "" and "" 10 cloverfield lane "" that have strong narratives. they enjoy complex storylines and unexpected twists, as seen in their rating of "" 10 cloverfield lane "". \newline
\textbf{suspense and thrills}: the user enjoys movies that are suspenseful and keep them on the edge of their seat. they appreciate claustrophobic settings, as seen in their rating of "" 10 cloverfield lane "". \newline
\textbf{action and adventure}: the user enjoys action - packed movies with a mix of adventure and espionage, as evident from their ratings of "" kingsman : the secret service "" and "" inception "". \newline
\textbf{language}: the user doesn't seem to have any issues with subtitles, as they didn't mention anything negative about the subtitles in their review of "" life of pi "". \newline
overall, the user enjoys complex, suspenseful storylines with strong performances, action - packed plots, and high - quality visuals. they are willing to pay a premium for blu - ray copies of movies and appreciate movies that keep them guessing until the end. \\
\midrule[0.3ex]
\end{tabular}}
\end{table*}

\begin{table*}[htb]\centering
\raggedright{\textbf{Llama-3-8B-it} $|$ $k=2$, Maximum Generation Length = 512}
\resizebox{\linewidth}{!}{
\begin{tabular}{p{1.1\linewidth}} 
\midrule[0.3ex]
Based on the user's ratings and reviews, here is a summary of their preferences : \newline
\textbf{movie genre}: the user enjoys action - packed movies, particularly sci - fi actioners ( inception ) and spy - thrillers ( kingsman : the secret service ). they also appreciate horror movies ( 10 cloverfield lane ) and adventure films ( life of pi ). \newline
\textbf{storytelling}: the user values complex, thought - provoking storylines with unexpected twists and turns. they enjoyed the suspenseful plot of 10 cloverfield lane and the mind - bending narrative of inception. \newline
\textbf{visuals}: the user appreciates visually stunning movies, noting the impressive visuals in life of pi. they also appreciated the claustrophobic settings and tense atmosphere in 10 cloverfield lane. \newline
\textbf{acting}: the user highly praised the acting in 10 cloverfield lane, noting that the performances were superb. they also enjoyed the ensemble cast in kingsman : the secret service. \newline
\textbf{pacing}: the user acknowledged that life of pi may have some slow parts, but overall, they enjoyed the film's epic journey. they appreciated the fast - paced action and suspenseful plot in 10 cloverfield lane and kingsman : the secret service. \newline
\textbf{language}: the user noted that life of pi had more subtitles than indicated, which may be a consideration for future purchases. \newline
\textbf{overall}: the user enjoys action - packed movies with complex storylines, suspenseful plots, and impressive visuals. they appreciate strong acting performances and are willing to overlook slower pacing parts in favor of enjoying the film's overall experience. \\
\midrule[0.3ex]
\end{tabular}}
\end{table*}

\begin{table*}[htb]\centering
\raggedright{\textbf{Llama-3-8B-it} $|$ $k=1$, Maximum Generation Length = 512}
\resizebox{\linewidth}{!}{
\begin{tabular}{p{1.1\linewidth}} 
\midrule[0.3ex]
Based on the user's ratings and reviews, here is a summary of their preferences:\newline
\textbf{Overall Preferences}: The user enjoys action - packed movies with suspenseful plots, thrilling endings, and strong performances. They appreciate visually stunning movies with great storytelling, cinematography, and direction. The user seems to be fond of sci - fi, adventure, and spy thrillers, as seen in the movies "" life of pi "", "" 10 cloverfield lane "", "" inception "", and "" kingsman : the secret service "". \newline
\textbf{Movie length and pacing}: The user mentioned that "" life of pi "" is a "" long movie "" and that some parts might be boring, suggesting that they prefer movies with a faster pace. However, they did not mention any issues with the pacing of the other movies, which may indicate that they are willing to invest time in a well - crafted story. \newline
\textbf{Characteristics of well-received movies} : The user enjoyed movies with well - executed suspense, surprise endings, and strong performances. They appreciated movies with great visuals, direction, and storytelling. The user seems to value movies that keep them guessing until the end, as seen in their praise for the unexpected ending of "" 10 cloverfield lane "". \newline
\textbf{Negative Feedback} : The user mentioned that "" life of pi "" has more subtitles than indicated, which might be a minor annoyance. They did not provide any negative feedback for the other movies, indicating that they were generally satisfied with their viewing experiences. \newline
\textbf{Budget and Price} : The user seems to be willing to pay a premium for high - quality movies, as evidenced by their purchases of blu - ray copies and willingness to pay upwards of \$ 10 for each movie.\\
\midrule[0.3ex]
\end{tabular}}
\end{table*}

\begin{table*}[htb]\centering
\raggedright{\textbf{Llama-3-8B-it} $|$ $k=0$, Maximum Generation Length = 512}
\resizebox{\linewidth}{!}{
\begin{tabular}{p{1.1\linewidth}} 
\midrule[0.3ex]
based on the user's reviews and ratings, here are some insights into their movie preferences : \newline
1. \textbf{story is important}: the user appreciates a good story, as evident from their reviews of "" life of pi "" and "" 10 cloverfield lane "". they mention the importance of a great story, message, and script. \newline
2. \textbf{action and suspense}: the user enjoys action - packed and suspenseful movies, as seen in their reviews of "" 10 cloverfield lane "", "" inception "", and "" kingsman : the secret service "". they appreciate films that keep them engaged and guessing until the end. \newline
3. \textbf{visually impressive}: the user is impressed by visually stunning movies, as evident from their review of "" life of pi "", which they praised for its great visuals. \newline
4. \textbf{credible acting}: the user values good acting, citing the superb acting in "" 10 cloverfield lane "" as one of the film's strengths. \newline
5. \textbf{frequent surprises}: the user seems to enjoy unexpected twists and surprises, as seen in their reviews of "" 10 cloverfield lane "" and "" inception "", where they were kept guessing until the end. \newline
6. \textbf{genre preferences}: the user appears to enjoy movies that fall under the action, adventure, horror, and sci - fi genres, as seen from their reviews of "" life of pi "", "" 10 cloverfield lane "", "" inception "", and "" kingsman : the secret service "". \newline
overall, the user seems to appreciate movies that have a strong narrative, exciting action sequences, and impressive visuals, with a focus on suspense, twists, and surprises. they value good acting and are open to trying different genres and themes.\\
\midrule[0.3ex]
\end{tabular}}
\end{table*}



\clearpage

\section{Interpretability Evaluation Details}
\label{app:survey_details}

\subsection{GPT-4 Study Details}

To mitigate the possible position bias and hallucination in GPT-4 evaluations~\citep{bai2022constitutionalaiharmlessnessai, chiang2023largelanguagemodelsalternative}, we randomly shuffle the order of two candidate movies and prompt GPT-4 to first give an explanation of the choice before outputting the final choice. 
The randomization ensures that any potential preference for the first or second option is distributed evenly across the dataset, effectively neutralizing systematic position-based influences. The explanation encourages the model to engage in a more thorough analysis of the given information before making a decision.

\subsection{Human Study Details}

Human participants ($n=150$, recruited via Amazon Mechanical Turk~\citep{Buhrmester2016}) are first presented with an informed consent form with a brief task description. We assess attentiveness using a screening task with three custom profile texts that are straightforward to answer. The main task asks participants to predict the next movies each of 9 users will watch based on their profile texts. Similar to the GPT-4 procedure, we show answer candidates in a shuffled order. Finally, we give a brief questionnaire about English language proficiency and movie domain knowledge. To maintain experimental integrity and balance, we implement a systematic approach to profile assignment. Each participant receives three profiles of each format in a randomized sequence. We ensure that no participant receives profiles from the same user, and all three profile formats generated for each user interaction history are distributed exactly once across participants. We give participants incentive to answer correctly by giving bonus payment for each correct answer.

To ensure high-quality responses, we implement multiple participant selection criteria, including residing in the U.S., a minimum 97\% prior task approval rating, and at least 1000 previously approved tasks. Compensation is structured to incentivize accuracy, with participants receiving a base payment of \$0.50 and an additional \$0.65 for each correct answer, allowing for a maximum earnings of \$6.35 for 3 screening questions, 9 main questions, and 2 demographics questions. Based on actual payments and recorded survey durations, the mean hourly rate for participants (within a z-score range of [-0.5, 0.5] for total survey duration) is \$20.35.

The distribution of task scores from post-screening participants was approximately normal. Most participants stated that they were fairly proficient in English and "moderately", "very", or "extremely" familiar with movies. We observed a slight trend where participants who self-described as more proficient in English took longer to answer each task - one potential cause for this is that MTurk who are less proficient (or are not answering in good faith) may prefer to submit random guesses after the screening quiz if they thought the profiles involved too much reading simply to reach the end of the task for payment. We also observed another slight trend where participants who declared themselves more familiar with movies to have lower inference accuracy.
The breakdown of average accuracy and average per-task duration across different demographic responses is shown in Tables \ref{tab:dem_english_effects} and \ref{tab:dem_moviefam_effects}.

{\renewcommand{\arraystretch}{1.0}
\begin{table}[htb]\centering
\resizebox{0.95\linewidth}{!}{\begin{tabular}[t]{cccccccc} 
\midrule[0.15ex]
"How well can you\\read English?" & \# Respondents & \multicolumn{3}{l}{Avg Accuracy} & \multicolumn{3}{l}{Avg Duration (sec)} \\
& & NoProfile-0 & \custompipelinename-0 & \custompipelinename & NoProfile-0 & \custompipelinename-0 & \custompipelinename \\
\midrule[0.05ex]
Not at all &   1 & 33.33\% &   0.00\% & 33.33\% &  0.95 &  1.85 &  2.78 \\
Not well   &  16 & 50.00\% & 100.00\% & 16.67\% &  2.94 &  2.95 &  2.63 \\
Well       &  39 & 58.12\% &  48.72\% & 64.10\% & 17.56 & 18.53 & 18.20 \\
Very well  & 108 & 57.10\% &  56.48\% & 58.02\% & 42.00 & 33.25 & 33.92 \\
(All)      & 150 & 57.11\% &  54.67\% & 58.89\% & 34.76 & 28.77 & 29.15 \\
\midrule[0.15ex]
\end{tabular}}
\caption{\textbf{English language responses for interpretability task, task accuracy, and average task duration. ($|z|>3$ outliers filtered out for duration.)}\label{tab:dem_english_effects}}
\vskip -0.3cm
\end{table}}

{\renewcommand{\arraystretch}{1.0}
\begin{table}[htb]\centering
\resizebox{0.95\linewidth}{!}{\begin{tabular}[t]{cccccccc} 
\midrule[0.15ex]
"How familiar are you\\with movies overall?" & \# Respondents & \multicolumn{3}{l}{Avg Accuracy} & \multicolumn{3}{l}{Avg Duration (sec)} \\
& & NoProfile-0 & \custompipelinename-0 & \custompipelinename & NoProfile-0 & \custompipelinename-0 & \custompipelinename \\
\midrule[0.05ex]
Not familiar        &   0 &       - &        - &       - &     - &     - &     - \\
Slightly familiar   &   0 &       - &        - &       - &     - &     - &     - \\
Moderately familiar &  16 & 60.42\% &  58.33\% & 64.58\% & 30.52 & 23.17 & 21.75 \\
Very familiar       &  69 & 57.49\% &  54.59\% & 58.45\% & 36.72 & 31.91 & 32.60 \\
Extremely familiar  &  65 & 55.90\% &  53.85\% & 57.95\% & 33.78 & 26.91 & 27.33 \\
(All)               & 150 & 57.11\% &  54.67\% & 58.89\% & 34.76 & 28.77 & 29.15 \\
\midrule[0.15ex]
\end{tabular}}
\caption{\textbf{Movie familiarity responses for interpretability task, task accuracy, and average task duration. ($|z|>3$ outliers filtered out for duration.)}\label{tab:dem_moviefam_effects}}
\vskip -0.3cm
\end{table}}

It is worth noting that many profiles in the \custompipelinename-0 baseline were more irregularly formatted than the other two conditions, which may contribute a large amount to this accuracy difference. Profile texts with missing line breaks or excess token spacing were much more common in the \custompipelinename-0 condition than in any other condition. We give examples of what this format difference may look like in Appendix \ref{app:example_gen}.

It is also important to highlight how we only compare \custompipelinename\ profiles against full item metadata with length 5. We speculate that if \custompipelinename\ were adapted to generate profiles for longer item interaction histories, the difference in interpretability would become much larger - the length and readability of a generated profile may stay relatively steady, while a full item history listing continues to grow longer and more difficult to read and manually synthesize as more items are added. Testing this will require improving \custompipelinename\ to accommodate longer input context lengths in some way.

We show examples of the main task texts below. The profile texts ("summary of their preferences" content) and candidate options are changed depending on condition and user interaction history, but the rest of the task template remains constant.
We also show the three main screening tasks for the human participants survey.

\begin{table*}[htb]\centering
\raggedright{\textbf{Human Evaluation Example}}
\resizebox{\linewidth}{!}{
\begin{tabular}{p{1.1\linewidth}} 
\midrule[0.3ex]
Here's some information about the user. \newline
\newline
--- \newline
Based on the user's ratings and reviews, here is a summary of their preferences: \newline

**Suspense and Drama**: The user enjoys Suspense and Drama movies, as evidenced by their positive ratings for "Allied" (4.4) and "Britannia" (4.0), both of which blur the lines between action and emotion. \newline

**Science Fiction**: The user is willing to engage with Science Fiction, but only if it meets their expectations. They were disappointed by "Arrival" (3.0), feeling misled by its genre-labeling, but enjoyed "Humanity's End" (4.0), praising its story and happy ending. \newline

**Storytelling**: The user values a good story, as seen in their appreciation for "Britannia" and "Humanity's End". They want to be entertained and enjoy a compelling narrative. \newline

**Action and Fantasy**: The user enjoys Action and Fantasy elements, as demonstrated by their positive ratings for "Britannia" and "Humanity's End", both of which feature action scenes and fantastical elements. \newline

**Visuals**: The user is willing to forgive some visual shortcomings, as evidenced by their appreciation for "Humanity's End" (2009), which had relatively dated CGI. \newline

**Happy Endings**: The user seems to appreciate happy endings, as indicated by their appreciation for the conclusion of "Humanity's End". \newline

**Budget-Friendly Options**: The user may be budget-conscious, as they rented "Arrival" instead of buying it, suggesting a willingness to try before committing to a purchase. \newline

Overall, the user enjoys Suspense and Drama, Science Fiction, Action, and Fantasy, prioritizing storytelling, happy endings, and budget-friendly options. \newline
--- \newline
\newline
The user watched one of the following two movies next. Which one do you think it is? \newline
\newline
Title: V for Vendetta \newline
Description: (average rating: 4.7) From Lilly and Lana Wachowski and Joel Silver, the masterminds behind The Matrix trilogy, comes another intriguing, action-packed fantasy-thriller. \newline
Category: Science Fiction, Suspense, Cerebral, Downbeat \newline
Price: 14.99 \newline
\newline
Title: 13 Hours: The Secret Soldiers of Benghazi \newline
Description: (average rating: 4.8) From director Michael Bay, "13 HOURS: THE SECRET SOLDIERS OF BENGHAZI" is the gripping true story of six elite ex-military operators assigned to protect the CIA who fought back against overwhelming odds when terrorists attacked a U.S. diplomatic compound on September 11, 2012. When everything went wrong, six men had the courage to do what was right. Based on the nonfiction book "13 Hours: The Inside Account of What Really Happened in Benghazi" by New York Times bestselling author Mitchell Zuckoff and Members of the Annex Security Team. \newline
Category: Movies \& TV, Paramount Home Entertainment, Action \& Adventure \newline
Price: 7.19 \\
\midrule[0.3ex]
\end{tabular}}
\end{table*}

\begin{table*}[htb]\centering
\raggedright{\textbf{GPT-4 Evaluation Example}}
\resizebox{\linewidth}{!}{
\begin{tabular}{p{1.1\linewidth}} 
\midrule[0.3ex]
Here's some information about the user.\newline\newline
Based on the ratings and reviews provided, here is a summary of the user's preferences:\newline
\textbf{Comedy}: The user may not enjoy comedies, as evidenced by their 2.0 rating for "Clue The Movie", which they found "not funny" and "missed the mark".\newline
\textbf{Christmas Movies}: The user may enjoy Christmas movies, as evidenced by their 5.0 rating for "Hallmark Hall of Fame: A Heavenly Christmas", which they found to be a heartwarming story about acceptance and self-love.\newline
\textbf{Dramas}: The user may appreciate dramas, as evidenced by their 5.0 rating for "Follow Me, Boys!", which they found to be a warm and humorous story.\newline
\textbf{Fantasy/SF Movies}: The user may enjoy fantasy/science fiction movies, as evidenced by their 5.0 rating for "Powder", which they found to be a powerful and uplifting story.\newline
\textbf{Storytelling}: The user may be sensitive to storytelling, as evidenced by their appreciation for the themes and messages presented in "Hallmark Hall of Fame: A Heavenly Christmas" and "Powder".\newline
\textbf{Pacing}: The user may be willing to sacrifice screen time for pacing, as evidenced by their suggestion to cut some scenes from "Follow Me, Boys!".\newline
\textbf{Story Simplicity}: The user may enjoy simpler storylines, as evidenced by their appreciation for the "seemingly more simple time" presented in "Follow Me, Boys!".\newline
Overall, the user may be drawn to heartwarming stories, uplifting themes, and powerful performances, while being sensitive to pacing and storyline complexity.\newline\newline
The user watched one of the following two movies next. Which one do you think it is?\newline\newline
A: \newline
Title: The Christmas Ornament\newline
Description: (average rating: 4.8) Widow Kathy Howard is having a hard time facing her first Christmas without her husband when she meets Christmas tree lot owner Tim Pierce who helps her to see that there is still hope and love in the world. Kathy however, is unsure if she is truly ready to move on. Will she stay stuck in her grief, or will Tim and the Christmas season help open her heart to a new life?\newline
Category: Movies \& TV, Holidays \& Seasonal, Christmas\newline
Price: 6.46\newline\newline
B: \newline
Title: Arachnophobia\newline
Description: (average rating: 4.7) Product Description Hollywood Pictures and Amblin Entertainment deliver the year's most electrifying big-screen roller coaster ride of a movie! Everyone is afraid of something ... for Dr. Ross Jennings (Jeff Daniels), his phobia is downright embarrassing. But when he moves his family to a small town, the one thing that bugs him most is now harming the townspeople at an alarming rate. For this unlikely hero, overcoming a childhood fear of spiders might just save the community, but it may already be too late! Directed by Frank Marshall (Executive Producer, BACK TO THE FUTURE, Producer, RAIDERS OF THE LOST ARK) and also starring John Goodman, this critically acclaimed breathtaking hit entertains with its terrific mix of thrills, chills, and laughter! Amazon.com Most horror movies depend on giant monsters; Arachnophobia gets just as many thrills out of creatures only a few inches long. A scientist (Julian Sands, Warlock , A Room with a View ) who's hunting a vicious new species of spider in Venezuela unknowingly ships one back to the U.S. It ends up in a small town where a new doctor (Jeff Daniels, The Purple Rose of Cairo , Something Wild ) is trying to establish a practice. When his patients start suddenly dying, Daniels suspects spiders--but no one takes him seriously because he's had a phobia about spiders since childhood. Arachnophobia builds a slow but relentless sense of menace and creepiness, mixed with a sneaky satire of small town life. If you're squeamish about spiders, this will get under your skin. Also featuring the ever-dependable John Goodman ( The Big Lebowski , Barton Fink ) as a comically zealous exterminator. --Bret Fetzer\newline
Category: Movies \& TV, Featured Categories, DVD, Action \& Adventure\newline
Price: 6.99\newline\newline
Instructions:\newline
FIRST provide a one-sentence justification of your choice. \newline
SECOND, on a new line, state only "A" or "B" to indicate your choice. Your response should use the format:\newline
Justification: <one-sentence explanation>\newline
Choice: <"A" or "B"> \\
\midrule[0.15ex]
Justification: Given the user's demonstrated preference for heartwarming, uplifting stories and Christmas themed movies, and their apparent distaste for comedies (which "Arachnophobia" blends with its horror), the user is likely to choose "The Christmas Ornament".\newline
Choice: A\\
\midrule[0.3ex]
\end{tabular}}
\end{table*}

\begin{table*}[htb]\centering
\raggedright{\textbf{Human Screening Task Question 1}}
\resizebox{\linewidth}{!}{
\begin{tabular}{p{1.1\linewidth}} 
\midrule[0.3ex]
Here's some information about the user. \newline
\newline
--- \newline
The user has watched these movies: \newline

Title: Star Wars Episode 2 \newline
Description: (average rating: 4.7) The STAR WARS saga continues on DVD with Episode II Attack of the Clones. Anakin Skywalker has grown into an accomplished Jedi apprentice, and he faces his most difficult challenge yet as he must choose between his Jedi duty and forbidden love. Relive the adventure the way it was meant to be seen in spectacular digital clarity, including the climactic Clone War battle and Jedi Master Yoda in the ultimate lightsaber duel. Experience this 2-disc set that features over six hours of bonus materials, and see how Episode II unlocks the secrets of the entire STAR WARS saga. \newline
Category: Movies \& TV, Featured Categories, DVD, Science Fiction \newline
Price: 8.49 \newline
Review from the user: (rating: 4.0) Just completing my collection. Nice box and format quality. Fun bonus media. \newline

Title: Star Wars Episode 3 \newline
Description: (average rating: 4.7) The Star Wars saga is now complete on DVD with Episode III REVENGE OF THE SITH. Torn between loyalty to his mentor, Obi-Wan Kenobi, and the seductive powers of the Sith, Anakin Skywalker ultimately turns his back on the Jedi, thus completing his journey to the dark side and his transformation into Darth Vader. Experience the breathtaking scope of the final chapter in spectacular clarity and relive all the epic battles including the final climactic lightsaber duel between Anakin and Obi-Wan. \newline
Category: Movies \& TV, Featured Categories, Science Fiction \newline
Price: 12.98 \newline
Review from the user: (rating: 4.0) Prequel trilogy, movie is OK, packaging is very nice \newline

Title: Westworld: The Complete First Season \newline
Description: (average rating: 4.6) Created for television by Jonathan Nolan and Lisa Joy, WESTWORLD is a dark odyssey about the dawn of artificial consciousness and the evolution of sin – exploring a world in which every human appetite, not matter how noble or depraved, can be indulged. Exploring what it means to be human through the eyes of the lifelike AI “hosts” in the park, the series investigates the boundaries of an exotic world set at the intersection of the near future and the reimagined past. Meticulously crafted and artfully designed. Westworld offers its guests an unparalleled, immersive world where they have the freedom to become who they’ve always wanted to be – or who they never knew they were. No rules, no laws, no judgment. Live without limits. Anthony Hopkins, Ed Harris, Evan Rachel Wood, James Marsden, Thandie Newton, Jeffrey Wright and more star in this one-hour drama series from J.J. Abrams Bad Robot Productions and Kilter Films. \newline
Category: Movies \& TV, Featured Categories, Drama, Westerns, Action \& Adventure \newline
Price: 12.99 \newline
Review from the user: (rating: 5.0) Really nice first season! \newline

Title: Star Wars Episode 4 \newline
Description: (average rating: 4.8) Young farm boy Luke Skywalker is thrust into a galaxy of adventure when he intercepts a distress call from the captive Princess Leia. The event launches him on a daring mission to rescue her from the clutches of Darth Vader and the Evil Empire. \newline
Category: Movies \& TV, Featured Categories, Science Fiction \newline
Price: 7.53 \newline
Review from the user: (rating: 5.0) finally the main trilogy! classic series. \newline
--- \newline
\newline
The user watched one of the following two movies next. Which one do you think it is? \newline
\newline
Title: Star Wars Episode 5 \newline
Description: (average rating: 4.8) The Rebels scatter after the Empire attacks their base on the ice planet Hoth. Han Solo and Princess Leia are pursued by Imperials, while Luke trains with Jedi Master Yoda. Luke battles Darth Vader and learns the shocking truth of his past. \newline
Category: Movies \& TV, Featured Categories, Science Fiction \newline
Price: 6.75 \newline
\newline
Title: Coco \newline
Description: (average rating: 4.8) Despite his family's baffling generations-old ban on music, Miguel dreams of becoming an accomplished musician like his idol, Ernesto de la Cruz. Desperate to prove his talent, Miguel finds himself in the stunning and colorful Land of the Dead following a mysterious chain of events. Along the way, he meets charming trickster Hector, and together, they set off on an extraordinary journey to unlock the real story behind Miguel's family history. \newline
Category: Movies \& TV, Featured Categories, DVD, Family \newline
Price: 9.96 \\
\midrule[0.3ex]
\end{tabular}}
\end{table*}

\begin{table*}[htb]\centering
\raggedright{\textbf{Human Screening Task Question 2}}
\resizebox{\linewidth}{!}{
\begin{tabular}{p{1.1\linewidth}} 
\midrule[0.3ex]
Here's some information about the user. \newline
\newline
--- \newline
Based on the ratings and reviews, it is clear that the user has a strong preference for animated films and themes of environmentalism, such as ""Nausicaä of the Valley of the Wind"" by Studio Ghibli. \newline
--- \newline
\newline
The user watched one of the following two movies next. Which one do you think it is? \newline
\newline
Title: Princess Mononoke \newline
Description: (average rating: 4.9) From the legendary Studio Ghibli, creators of Spirited Away, and Academy Award®-winning director Hayao Miyazaki, comes an epic masterpiece that has dazzled audiences worldwide with its breathtaking imagination, exhilarating battles, and deep humanity. Inflicted with a deadly curse, the young warrior Ashitaka heads west in search of a cure. There, he stumbles into a bitter conflict between Lady Eboshi, the proud people of Iron Town, and the enigmatic Princess Mononoke, a young girl raised by wolves, who will stop at nothing to prevent the humans from destroying her home and the forest spirits and animal gods who live there. \newline
Category: Movies \& TV, Featured Categories, DVD, Kids \& Family, Anime \& Manga \newline
Price: 13.99 \newline
\newline
Title: Back to the Future \newline
Description: (average rating: 4.8) From Academy Award®-winning filmmakers Steven Spielberg and Robert Zemeckis comes Back to the Future -- the original, ground-breaking adventure that sparked one of the most successful trilogies ever. When teenager Marty McFly (Michael J. Fox) is blasted to 1955 in the DeLorean time machine created by the eccentric Doc Brown (Christopher Lloyd), he finds himself mixed up in a time-shattering chain reaction that could vaporize his future -- and leave him trapped in the past. Powered by innovative special effects, unforgettable songs and non-stop action, Back to the Future is an unrivaled adventure that stands the test of time! \newline
Category: Movies \& TV, Science Fiction \& Fantasy, Kids \& Family, Comedy \newline
Price: 8.99 \\
\midrule[0.3ex]
\end{tabular}}
\end{table*}

\begin{table*}[htb]\centering
\raggedright{\textbf{Human Screening Task Question 3}}
\resizebox{\linewidth}{!}{
\begin{tabular}{p{1.1\linewidth}} 
\midrule[0.3ex]
Here's some information about the user. \newline
\newline
--- \newline
This viewer appreciates films that challenge conventions, combining emotional depth with unique storytelling. \newline
They enjoy visually compelling films like Inception, especially those with layered narratives and personal stakes. \newline
Their enthusiasm for the absurd yet heartfelt Swiss Army Man suggests a preference for unconventional, genre-blending films that evoke strong emotional reactions. \newline
With Scott Pilgrim vs. the World, they appreciate pop culture-infused storytelling, even if they don't grasp every reference, valuing the film's energetic presentation over strict narrative coherence. In their review, they mention how they are more familiar with Chinese immigrant culture and would like to see greater representation of that in film. \newline
Overall, they gravitate toward films that provoke thought and emotion through distinct, imaginative approaches. \newline
--- \newline
\newline
The user watched one of the following two movies next. Which one do you think it is? \newline
\newline
Title: Everything Everywhere All at Once \newline
Description: (average rating: 4.6) A middle-aged Chinese immigrant (Michelle Yeoh) is swept up into an insane adventure in which she alone can save existence by exploring other universes and connecting with the lives she could have led. \newline
Category: Movies \& TV, Featured Categories, DVD, Science Fiction \newline
Price: 10.49 \newline
\newline
Title: Alita: Battle Angel \newline
Description: (average rating: 4.6) A cyborg uses her prodigious fighting skills to take on corrupt authorities in a future dystopia. \newline
Category: Movies \& TV, DVD, Action \& Adventure \newline
Price: 7.74 \\
\midrule[0.3ex]
\end{tabular}}
\end{table*}



\end{document}